\documentclass[usenatbib]{mn2e}
\usepackage{amsmath}
\usepackage{txfonts}
\usepackage{graphicx}
\usepackage{array}

\newcommand{\be}{\begin{equation}}
\newcommand{\ee}{\end{equation}}
\newcommand{\bea}{\begin{eqnarray}}
\newcommand{\eea}{\end{eqnarray}}

\newcommand{\oiii}{[O\,{\sc iii}]}
\newcommand{\ovi}{[O\,{\sc vi}]}
\newcommand{\civ}{C\,{\sc iv}}

\newcommand{\ci}{C\,{\sc i}}
\newcommand{\cifull}{C\,{\sc i} ($^3$P$_2  \rightarrow^3$P$_1$)}
\newcommand{\cifullio}{C\,{\sc i} ($^3$P$_1  \rightarrow^3$P$_0$)}

\newcommand{\co}{CO\,{\sc (1$\rightarrow$0) }}
\newcommand{\coiii}{CO\,{\sc (3$\rightarrow$2) }}

\newcommand{\hst}{{\sl HST} }

\newcommand{\evla}{{\sl JVLA } }
\newcommand{\jvla}{{\sl JVLA } }

\newcommand{\gbt}{{\sl GBT } }
\newcommand{\alma}{{\sl ALMA}}

\newcommand{\mnras}{MNRAS}

\newcommand{\apj}{ApJ}
\newcommand{\apjl}{ApJL}
\newcommand{\aap}{A\&A}

\newcommand{\prd}{Phys. Rev. D}

\title[Preferential magnification in IRAS~FSC10214+4724  \--- II.]{\mbox{The preferentially magnified active nucleus in IRAS\,F10214+4724\,--\,II.} Spatially resolved cold molecular gas}

\author[Deane et al.]{R.P. Deane$^{1,2}$\thanks{E-mail: roger.deane@astro.ox.ac.uk}, I. Heywood$^1$, S. Rawlings$^1$, P.J. Marshall$^1$
\vspace*{3pt}\\
\noindent $^1$Astrophysics, Department of Physics, University of Oxford, Keble Road, Oxford, OX1 3RH, UK \\
\noindent $^2$Astrophysics, Cosmology and Gravity Centre, Department of Astronomy, University of Cape Town, Private Bag X3, Rondebosch 7701, South Africa}

\begin{document}

\date{Accepted 2013 May 31}

\pagerange{\pageref{firstpage}--\pageref{lastpage}} \pubyear{2013}

\maketitle

\label{firstpage}

\begin{abstract}We present Jansky Very Large Array (JVLA) observations of the cold (CO\,{\sc (1$\rightarrow$0)}) molecular gas in IRAS\,F10214+4724, a lensed Ultra-Luminous InfraRed Galaxy (ULIRG) at $z=2.3$ with an obscured active nucleus. The galaxy is spatially and spectrally well-resolved in the \co emission line. The total intensity and velocity maps reveal a reasonably ordered system, however there is some evidence for minor merger activity. A \co counter-image is detected at the \mbox{3-$\sigma$}~level. Five of the 42~km\,s$^{-1}$ channels (with $>$5-$\sigma$ detections) are mapped back into the source plane and their total magnification posterior probability distribution functions (PDFs) are sampled. This reveals a roughly linear arrangement, tentatively a rotating disk. We derive a molecular gas mass of $M_{\rm gas} = 1.2 \pm 0.2 \times 10^{10}$~M$_{\odot}$, assuming a ULIRG $L_{\rm CO}$-to-$M_{\rm gas}$ conversion ratio of $\alpha = 0.8$~M$_{\odot}$ (K\,km\,s$^{-1}$\,pc$^2$)$^{-1}$ that agrees well with the derived range of $\alpha = 0.3 - 1.3$~M$_{\odot}$~(K\,km\,s$^{-1}$\,pc$^2$)$^{-1}$ for separate dynamical mass estimates at assumed inclinations of $i = 90^\circ - 30^\circ$. The lens modelling and \co spectrum asymmetry suggest that there may be substantial (factor $\sim$2) preferential lensing of certain individual channels, however the \co spatially-integrated channel flux uncertainties limit the significance of this result. Based on the AGN and \co peak emission positions and the lens model, we predict a distortion of the CO Spectral Line Energy Distribution (SLED) where higher order ${\sl J}$ lines that may be partially excited by AGN heating will be preferentially lensed owing to their smaller solid angles and closer proximity to the AGN and therefore the cusp of the caustic. Comparison with other lensing inversion results shows that the narrow line region and AGN radio core in IRAS\,F10214+4724 are preferentially lensed by a factor $\gtrsim3$ and 11 respectively, relative to the molecular gas emission. This distorts the global continuum emission Spectral Energy Distribution (SED) and strongly suggests caution in unsophisticated uses of IRAS\,F10214+4724 as an archetype high-redshift ULIRG. We explore two Large Velocity Gradient (LVG) models, incorporating spatial \co and \coiii information and present tentative evidence for an extended, low excitation cold gas component that implies that the total molecular gas mass in IRAS\,F10214+4724 is a factor $\gtrsim$2 greater than that calculated using spatially unresolved CO observations.

\end{abstract}


\section{Introduction}\label{section_introduction}

Since the early 1990s it has been clear that high-redshift ultra-luminous infrared galaxies (ULIRGs) represent a cosmologically important population. The prodigious star formation rates in these objects at early cosmic epochs suggest they are the likely progenitors of today's most massive elliptical galaxies. Statistical analyses of these submillimetre-bright galaxies (SMGs and ULIRGs) show that they host the bulk of star formation at high redshift \citep[e.g.][]{Hughes1998,Chapman2005}, providing strong evidence for the cosmic down-sizing model where mass assembly, black hole growth and star formation takes place in the most massive systems first, contrary to a bottom-up hierarchical buildup \citep[e.g.][]{Bundy2005,Hasinger2005}. Despite its fundamental importance, a detailed study of this population remains challenging for a variety of reasons including cosmic dimming, dust extinction, instrument sensitivity and spatial resolution. 

We have undertaken an extensive multi-wavelength study of one of the earliest detected high-redshift ULIRGs, IRAS\,F10214+4724 (IRAS~10214 hereafter). In this paper we explore the cold gas properties as probed by the \co emission line. The {\sl Jansky VLA} and the {\sl IRAM Plateau de Bure Interferometer} have undergone a remarkable improvement in sensitivity at millimetre and centimetre wavelengths. This combined with the larger bandwidths has enabled more efficient, dedicated studies of high-redshift galaxies though the various transitions of the CO ladder \citep[e.g.][]{Greve2005,Tacconi2008,Carilli2010,Daddi2010,Ivison2011}. IRAS~10214 itself was the first high-redshift galaxy detected in CO \citep{Brown1991,Solomon1992}, however, this and most other high-redshift detections that followed were made in the mid- to high-{\sl J} lines, primarily due to the receiver sensitivity and available atmospheric windows. Relative to the rest of the CO ladder, the {\sl J} 1$\rightarrow$0 line is a less biased tracer of the total molecular mass, since higher rotational lines require higher temperatures and densities for excitation. Some observations have shown evidence for extended gas reservoirs at low excitation temperatures \citep[see][]{Carilli2010,Riechers2011SMGs,Ivison2011}, suggesting that the mid-to-high-{\sl J} lines were biased tracers of the total molecular gas content in higher redshift galaxies. However, \citet{Riechers2011} performed a \co analysis of five high-redshift lensed quasars (including IRAS 10214) and found no evidence for gas reservoirs untraced by the previously observed higher-{\sl J} CO lines, based on single temperature LVG models. However, this analysis was based on unresolved (or barely resolved) \co imaging and the assumption that all physical components undergo the same magnification boost. Here we wish to address the following question: do the AGN and star forming components undergo significantly different magnification boosts in IRAS 10214, by virtue of their different size and centroid positions? This is essentially a case study of the effective chromacity\footnote{This is caused by different emission regions undergoing differing magnification boosts due to their relative size and position with respect to the gravitational lens caustic.} of strongly-lensed systems selected at mid-infrared wavelengths. This of course applies to both the continuum SED
as well as the CO Spectral Line Energy Distribution (SLED). It is the second in a series and builds on the lens modelling and multi-wavelength investigation of the hidden quasar in IRAS~10214 in \citet[][{\bf D13a} hereafter]{Deane2013a}.

This paper is structured as follows: in \S2 we describe the \evla observations; in \S3 we present the spatially resolved \co maps (channel, total intensity and velocity). \S4 investigates the source plane parameters of these maps, deriving the intrinsic physical properties. In \S5 we synthesise these new observations with the multi-wavelength view of this galaxy and close with conclusions in \S6. Throughout this paper we assume a concordance cosmology of $\Omega_{\rm M}$ = 0.27, $\Omega_{\Lambda}$ = 0.73, and $H_0$ = 71 km\,s$^{-1}$\,Mpc$^{-1}$ \citep{Spergel2007}, which yields an angular size scale of 8.3~kpc\,arcsec$^{-1}$ at the redshift of IRAS 10214 ($z$ = 2.2856, \citealt{Ao2008}). We use the radio convention and define velocity $V = c(\nu_{\rm line} - \nu)/\nu_{\rm line}$, so quoted velocities are intrinsic and negative if blue-shifted with respect to the redshift of the IRAS 10214.

\section{Observations}

We observed the \co rotational line, redshifted to $\nu_{\rm obs}$~=~35.08376~GHz, toward IRAS 10214 using the \evla (C~configuration) in October 2010. Total on-source integration time was 3~h. The WIDAR correlator was used in OSRO-2 spectral line mode, employing 128 MHz of bandwidth per polarisation, split into 256 channels. After applying a natural weighting scheme to the {\sl uv}-data, a 500 kHz channel sensitivity of $\sim$500 $\mu$Jy\,beam$^{-1}$ is achieved. Absolute flux calibration was performed with the quasar {\sl J}\,0713+4349, while phase and bandpass calibration were carried out with the nearby ($\Delta \theta  = 1^{\circ}$) {\sl VLBA} calibrator, {\sl J}\,1027+4803. The target was observed for 2.3 minutes in each 3 minute phase-target cycle to track the rapid phase variations at this frequency. 

All data reduction was performed using standard reduction techniques in {\sc aips}, however, phase delay corrections were required and were calculated with the {\sc fring} task. Additional data inspection was performed with {\sc casa}. Data were edited in the full spectral and temporal resolution of 500 kHz and 1 second respectively. After phase calibration, the data were averaged into 5 MHz ($\Delta V = 42$ km\,s$^{-1}$) channels for imaging. All maps were generated with natural weighting yielding a synthesised beam size $\theta_{\rm syn} \sim 0.82 \times 0.62 \ \mathrm{arcsec}^2$.

\section{Results}

\begin{figure}
\includegraphics[width=0.47\textwidth]{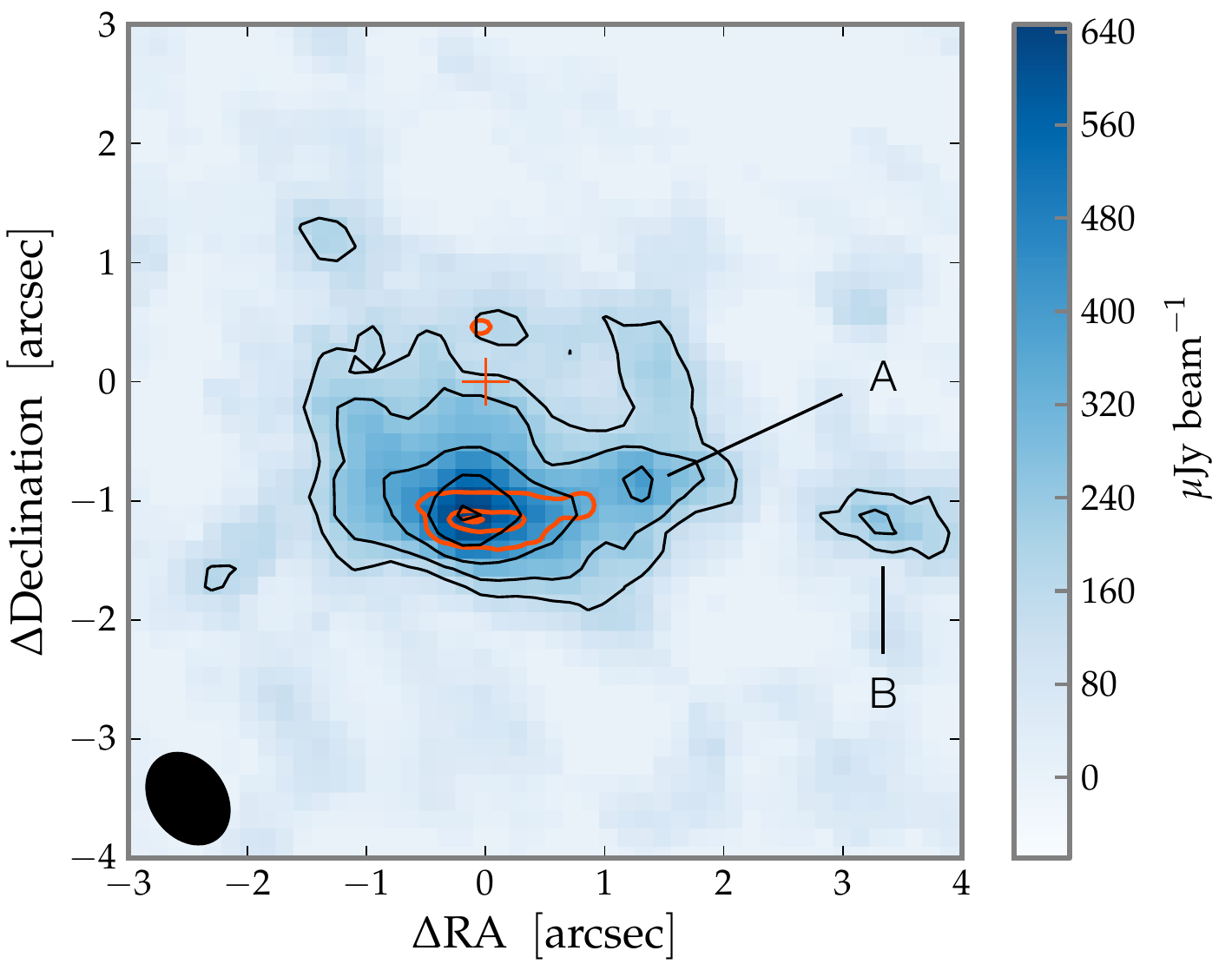}
\caption{\co intensity map of IRAS~10214 integrated over eight 42 km\,s$^{-1}$ channels. \hst rest-frame ultraviolet ($\lambda_{\rm obs} \sim 0.8~\mu$m) contours are overlaid in red and the centroid of the lens galaxy is indicated with a red cross. The \co contours (black) are at 2-$\sigma$ intervals and start at -3, 3-$\sigma$ where $\sigma$ = 60~$\mu$Jy\,beam$^{-1}$. The synthesised beam of 0.82$\times$0.62 arcsec$^2$ is shown in the bottom left corner. The map co-ordinates are centred on the lensing galaxy \hst\,F160W centroid (RA = 10$^{\rm h}$ 24$^{\rm m}$ 34.5622$^{\rm s}$, Dec = 47$^{\circ}$ 09' 10.809'', see {\bf D13a}). }
\label{fig:COmom0}
\end{figure}

\begin{figure*}
\includegraphics[width=0.9\textwidth]{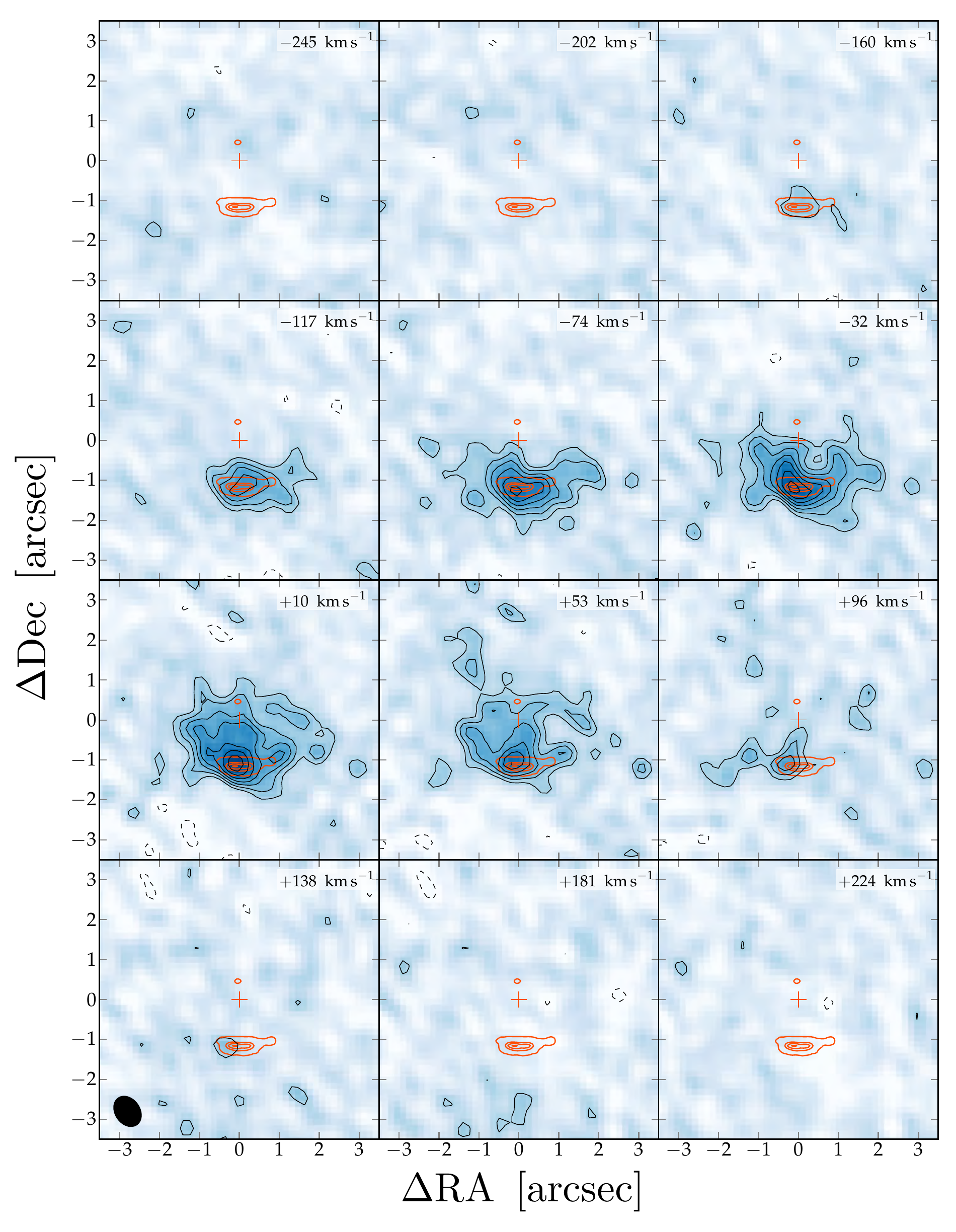}
\caption{\co channel maps of IRAS~10214. Contours are at 1-$\sigma$ intervals and start at -2.5, 2.5-$\sigma$ where $\sigma$ = 0.17 mJy\,beam$^{-1}$.  \hst rest-frame ultraviolet contours are overlaid in red and the centroid of the lens galaxy is indicated with a red cross. The synthesised beam of 0.82$\times$0.62 arcsec$^2$ is shown in the bottom left corner. Note that these channels are entirely independent (i.e. no form of Hanning smoothing was applied). The separate CO `core' we propose in \S\ref{sec:reservoir} is most evident at velocities $V = +10 \ {\rm and} \ +53 \, {\rm km\,s}^{-1}$, therefore spanning $>80  \, {\rm km\,s}^{-1}$. }
\label{fig:chanMaps}
\end{figure*}

The integrated intensity map (Fig.~\ref{fig:COmom0}) was generated by clipping all emission below 2-$\sigma$ in individual channels, before their summation. The map reveals a clearly extended morphology, particularly in the east-west direction which spans $\sim 3.5$~arcsec. An `arc-like' structure is observed, and this is clearer still in a few of the channel maps. The peak of the \co total intensity map is co-spatial with the {\sl Hubble Space Telescope (\hst)} rest-frame ultraviolet map (F814W filter, centred on 0.8~$\mu$m).

In Fig.~\ref{fig:chanMaps} we show twelve 42 km\,s$^{-1}$ channel maps of IRAS~10214, each with a noise level of $\sigma \sim 170 \ \mu$Jy\,beam$^{-1}$. There is clear emission above 3-$\sigma$ significance associated with the \hst\,F814W maps (rest-frame ultraviolet; red contours) for a velocity range of $\sim$340~km\,s$^{-1}$. These channels are entirely independent (i.e. no Hanning smoothing is applied after channel averaging).

As discussed in both {\bf D13a} and {\bf D13c}, the radio and optical reference frames are checked for any systematic offsets by \citet{Lawrence1993} who compare the optical and radio positions of 20 compact radio sources in the region around IRAS~10214 and find no significant mean difference. Since we use the same data, and our calibrators (as well as the 8~GHz centroids) are consistent, we assume we have an equivalent astrometric matching between radio and optical reference frames. However, they note the error on the mean difference is 0.2~arcsec which is taken as the systematic uncertainty. 

Accounting for the lower spatial resolution of this map, the source is significantly more extended in \co than the optical, near-infrared and radio components presented in {\bf D13a}. There is a second peak toward the west (labelled A in Fig.~\ref{fig:COmom0}), either suggestive of very clumpy structure or of minor merger activity not seen in other observed parts of the spectrum (which are spatially resolved). There is an additional component (labelled B) $\sim3.5$~arcsec west of the CO peak. This appears to be real feature as it is discernible at the 2-3-$\sigma$ level in three of the 42 km\,s$^{-1}$ channels. More tentative is an apparent 3-$\sigma$ detection of the CO counter-image, roughly co-spatial with the \hst\,F814W counter-image. Although highly uncertain, a naive arc-to-counter-image flux ratio $\check{\mu} = 7 \pm^6_2$ is calculated. This is done by fitting a point spread function (PSF) to the counter-image and an elongated 2D Gaussian to the arc. The quoted uncertainty is the quadrature sum of the fitting error and flux uncertainty. As we will derive in \S\ref{sec:srcplane}, the most probable model (generated with the mean of each parameter's PDF) has a counter-image that is co-spatial with the \hst\,F814W counter-image and not slightly west-ward as seen in the \co total intensity map. However, this $\sim$100 mas offset is within the positional uncertainty of the counter-image ($\sigma_{\theta} = 0.5\,{\rm FWHM}/\,$S/N$\sim 130$~mas). 

\begin{figure}
\includegraphics[width=0.47\textwidth]{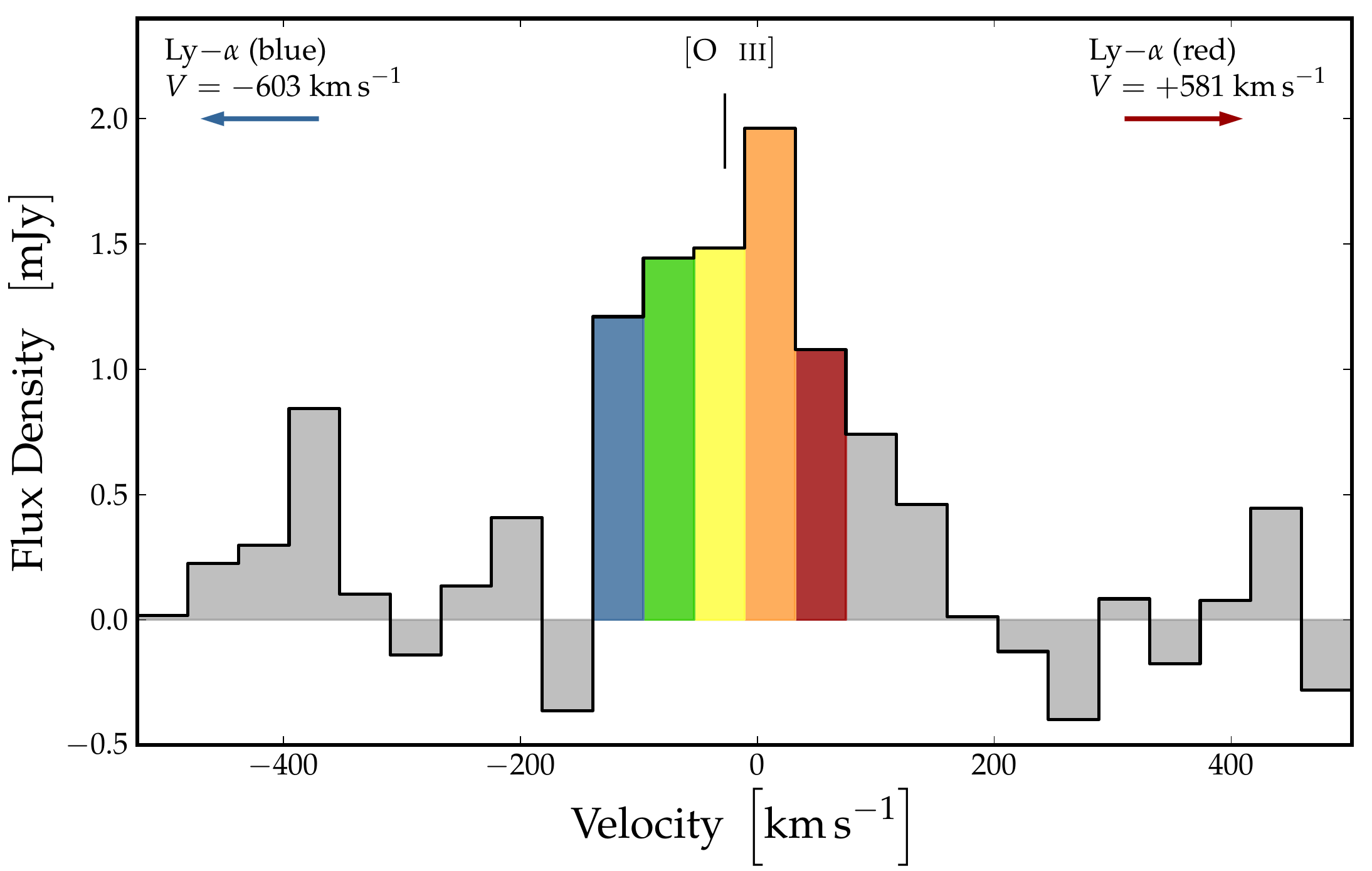} 
\caption{Spectrum of IRAS~10214 showing 24 $\times$ 42 km\,s$^{-1}$ channels covering $\sim$1000 km\,s$^{-1}$.  The asymmetric profile is also seen in the \coiii and C\,{\sc i} spectra reported in \citet{Ao2008}, however the spatially-integrated channel noise is $\sim 440\, \mu$Jy\,beam$^{-1}$ limiting the significance of this result. A more detailed discussion of the {\sl Ly-}$\alpha$ and other spectral line velocity offsets is presented in \S\ref{sec:redshifts} and summarised in Table~\ref{tab:redshifts}.  }
\label{fig:spec}
\end{figure}

The integrated spectrum shown in Fig.~\ref{fig:spec} has a Gaussian-fitted mean redshift of $z=2.28555$ $\pm$0.00005 (this excludes the Doppler tracking systematic uncertainty). We use a Markov Chain Monte Carlo (MCMC) algorithm to fit a Gaussian to the integrated spectrum in order to determine a realistic estimate of the uncertainty, and propagate these to any derived physical properties. The result is a peak intensity $S_\nu = 1.73 \pm 0.28$~mJy; velocity FWHM $V_{\rm FWHM} = 194 \pm 31$~km\,s$^{-1}$; and mean velocity $V_{\rm mean} = -14 \pm 13$~km\,s$^{-1}$. The mean velocity assumes the systemic redshift is $z = 2.2856$. If the mean velocity is fixed to zero, then there is a negligible change ($< 1$~percent) to $S_\nu$ and $V_{\rm FWHM}$. The velocity integrated \co flux density is $I_{\rm CO(1-0)} = 358 \pm 80$~mJy\,km\,s$^{-1}$, where the uncertainty is the quadrature sum of the MCMC-derived Gaussian amplitude and FWHM uncertainty. The velocity-integrated \co flux density is consistent with the \citet{Riechers2011} \gbt observation ($I_{\rm CO(1-0)} = 337 \pm 45$~mJy\,km\,s$^{-1}$); and the same authors \jvla D-configuration observation ($I_{\rm CO(1-0)} = 434 \pm 47$~mJy\,km\,s$^{-1}$). 

Our values imply $L'_{\rm CO} = 9.2 \pm 2.0 \times 10^{10}$~K\,km\,s$^{-1} \ \mu_{\rm CO(1-0)}^{-1}$, where $\mu_{\rm CO(1-0)}^{-1}$ is the magnification of the \co emitting gas. The spectrum appears asymmetric with a more prominent blue-wing (negative velocities), similar to the \coiii and C\,{\sc i} spectra reported in \citet{Ao2008}. This is consistent with \citet{Ao2008} who report a similar asymmetry in the C\,{\sc i} and \coiii integrated spectra, but not in the higher-order CO lines. However, the spatially-integrated \co channel noise is $\sim 440\, \mu$Jy\,beam$^{-1}$ and so the apparent asymmetry does not have high significance. We do not detect any continuum emission to a 3-$\sigma$ level of 90~$\mu$Jy\,beam$^{-1}$. We select channels that are above 5-$\sigma$ to perform a lensing analysis, which results in a total of 5 channels to investigate in the source plane.

\begin{figure}
\includegraphics[width=0.47\textwidth]{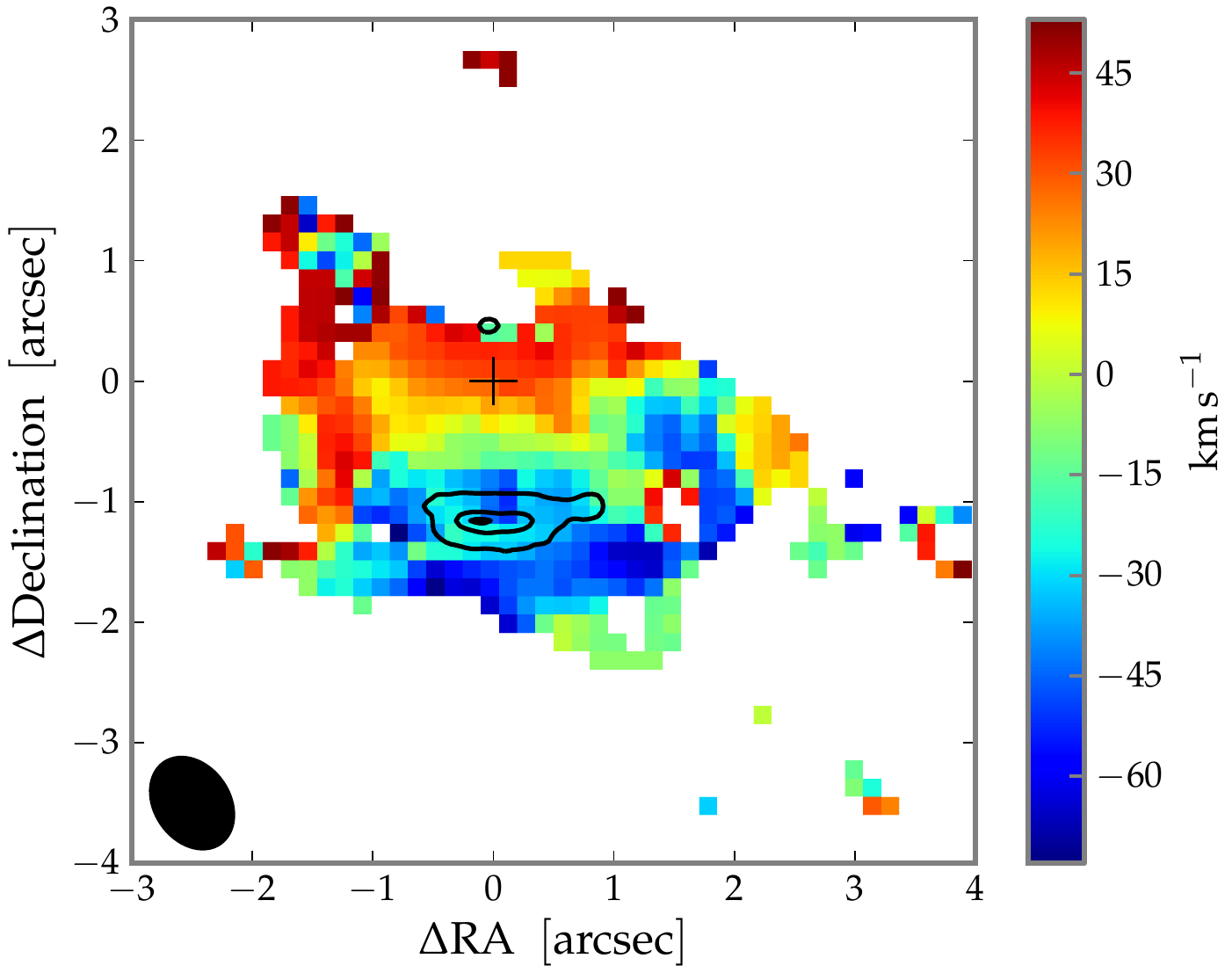} 
\caption{\co intensity-weighted velocity map of IRAS~10214, revealing a north-south velocity gradient. White pixels represent values outside the range shown by the colourbar. The white pixels within the velocity field in the south-west are roughly co-spatial with the secondary peak (labelled {\sl A} in Fig.~\ref{fig:COmom0}) and at a velocity of $V \sim$100 km\,s$^{-1}$, however are not included here to better illustrate the velocity field. \hst rest-frame ultraviolet contours are overlaid in black and the centroid of the lens galaxy is indicated with a cross. }
\label{fig:COmom1}
\end{figure}

The intensity-weighted mean, image-plane velocity map is shown in Fig.~\ref{fig:COmom1}. This map was is generated by applying a 2-$\sigma$ clip to the four channels with highest flux density in the integrated spectrum (Fig~\ref{fig:spec}). The velocity field spans a range $\Delta V = 168$~km\,s$^{-1}$ and reveals a clear north-south gradient (or NNE-SSW). Velocities outside this range are clipped and shown in white. \citet{Ao2008} mapped the \coiii line towards IRAS~10214 and found a NE-SW velocity gradient. The same authors find an east-west velocity gradient C\,{\sc i} ($^3$P$_2  \rightarrow^3$P$_1$), however IRAS~10214 was barely resolved in the north-south direction making a direct comparison difficult. Moreover, their velocity range covered $\sim 90$~km\,s$^{-1}$, half of what is traced by our \co map. Finally, the apparent second \co peak appears to be at a higher recession velocity, as seen in the channel maps.

\section{Source Plane Inversion}\label{sec:srcplane}

\begin{figure*}
\includegraphics[width=1\textwidth]{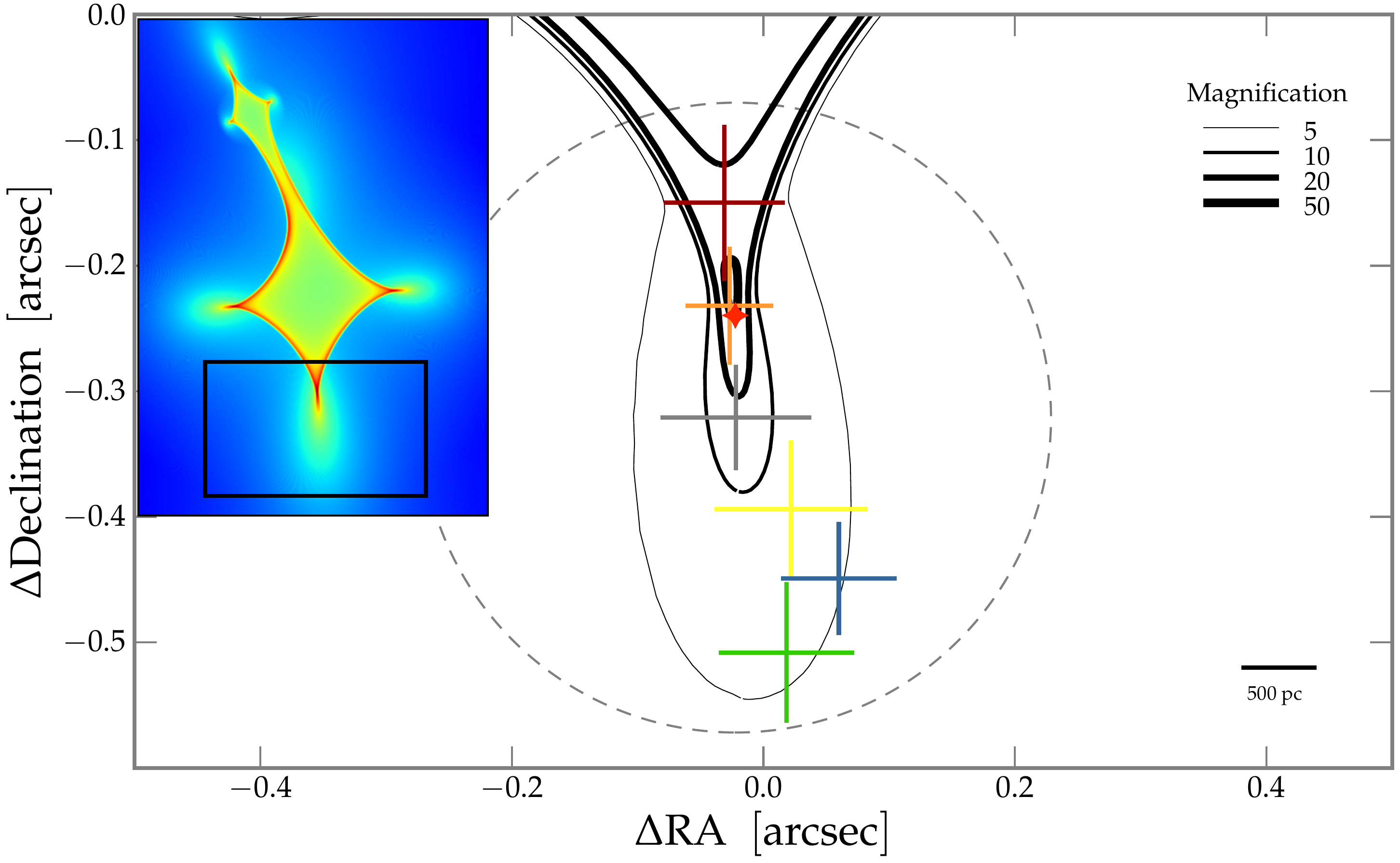} 
\caption{Source plane reconstruction showing the centroids of individual channel maps that have an integrated flux density $>$5-$\sigma$ ($\sim$1 mJy\,beam$^{-1}$ in a 42 km\,s$^{-1}$ channel). The cross sizes indicate the MCMC-derived 68~percent confidence level of the source-plane centroids. The crosses are colour-coded by velocity: blue representing the most blue-shifted channel ($V = -117$ km\,s$^{-1}$) and increasing in steps of $\Delta V = 42$ km\,s$^{-1}$. The grey cross and corresponding grey dashed circle represent the best-fit model to the total intensity map (Fig~\ref{fig:COmom0}), where the scale radius is $r_s = 5.8\pm^{0.9}_{0.7}$~kpc. The VLBI-detected radio core position is indicated by the red diamond which is consistent with the vector linking all CO centroids and is co-spatial with the \co channel closest to zero velocity (orange, +10~km\,s$^{-1}$). Note that the \co size estimate is based on the integrated flux density map, which is a combination of eight 42~km\,s$^{-1}$ channels ($\Delta V = 336$~km\,s$^{-1}$), and so should be considered as a lower limit. The black lines represent contours of magnification, based on the lens model derived in {\bf D13a}.}
\label{fig:srcPlane}
\end{figure*}

We estimate the magnification of the \co channel maps and total intensity map using a Bayesian MCMC algorithm, previously presented in {\bf D13a}. For the purposes of this work, we fix the lens model parameters to those defined in Lens Model~A in {\bf D13a}, which was derived using a deep, rest-frame ultraviolet \hst map (F814W filter). See {\bf D13a} for further details on the derivation of this lens model. We derive the source-plane \co properties, given the data which are the individual channel map pixel values presented in Fig.~\ref{fig:chanMaps}. We include channels which have an integrated flux density with $>$5-$\sigma$ significance. The PSF, which is well-defined in radio interferometric observations, has major and minor axis FWHM of 0.82$\times$0.62~arcsec$^2$ at a position angle of 34.3$^\circ$ east of north. We assume spherically symmetric, Gaussian components for the \co source model. A Gaussian profile is a highly simplified representation of high-redshift gas morphology, which observations and simulations show to be dominated by large star-forming clumps. However, this simplification has two clear advantages: (1) the lensing inversion computational requirements are drastically decreased; (2) the resultant models allow a uniform comparison (specifically of the centroid and radius uncertainties) with other \co channels as well as the source-plane properties of other radio and optical maps reported in {\bf D13a}. We make the assumption that the \co channel map noise properties are uncorrelated which is not true. However, this is unlikely to make a large impact since each channel map has its dominant peak and arc structure reasonably reproduced, with residuals that appear to be dominated by the simplification in source-plane structure. Proper treatment of the correlated noise would be important in a pixel-by-pixel reproduction of the source-plane map, however at this low S/N the benefits of such an approach are not immediately clear. A good example of an alternative approach is that performed in \citet{Riechers2008}, who model 7 independent CO (2-1) channel maps of the molecular Einstein ring in PSS~{\sl J}\,2322+1944 using a pixel-based source-plane reconstruction. Their resulting lensing inversion does not allow the source-plane structure uncertainties to be easily described, however it does reproduce the image-plane structure very well. 

With the lens model fixed, we invert the five image-plane channels as well as the integrated intensity map (average of eight channels) into the source plane. We vary the source-plane centroid coordinates and scale radius which are all assigned uniform priors. MCMC chains are run in the same manner as described in {\bf D13a}: a Metropolis-Hastings algorithm (with a Gaussian proposal distribution) is tuned to have an acceptance rate of 20~percent. We find convergence occurs after $\sim$10$^5$ iterations, however this is significantly shortened if the eigenvectors and eigenvalues are estimated (from a $> 10^{5}$ iteration run) and used to transform the parameter vector into uncorrelated space where proposal distributions are more efficient in achieving convergence. Our MCMC algorithm samples the unnormalised posterior PDF of the centroid $RA, Dec$ and scale radius $r_s$, quantifying in the level of uncertainty of each free parameter.

\subsection{Channel Centroid Positions}

In Fig.~\ref{fig:srcPlane} we plot the resultant centroid estimates of the lens inversion process. The black lines represent contours of magnification around the caustic (see legend, top right). The coloured crosses represent the mean of the centroid posterior PDFs, with the 68~percent confidence level indicated by the cross length. The colour is coded by recession velocity, starting with  the most blue-shifted channel ($V = -117$~km\,s$^{-1}$) and increasing in steps of $\Delta V = 42$ km\,s$^{-1}$. This reveals a roughly linear source-plane structure, suggesting regular rotation in the disk and consistent with the velocity field in Fig.~\ref{fig:COmom1}. We note that these two observational results are consistent with, but not evidence for, a regularly rotating molecular gas disk. A straight line fit to the channel source-plane centroids results in a vector with a position angle $PA_{\rm CO,vel} = 16^{\circ} \pm 7^{\circ}$ east of north. We also plot the total intensity map source-plane centroid with a black cross. The dashed black circle represents the scale radius posterior PDF mean of the integrated intensity map which is consistent with all the channel map source-plane radii ($r_s = 5.8\pm^{0.9}_{0.7}$~kpc). This compares very well to the intrinsic size of PSS~{\sl J}\,2322+1944 ($r_{\rm eff} \sim$5~kpc), which also has a similar gas mass ($M_{\rm gas} \sim 2\times10^{10}$~M$_\odot$) and shows evidence for a regularly rotating disk \citep{Riechers2008}. However, the authors suggest that the system is likely to be interacting based on the apparent angular offset between the molecular gas and optical emission peaks. 

The north-south velocity gradient and the obscured AGN properties of this galaxy naively imply that a putative molecular gas disk would have a position angle of $\sim 0^\circ$ east of north. However, the image plane velocity field has a near unity axis ratio. Based on the source plane positions of the individual channels, most of the magnification is in the east-west direction, perpendicular to the expected `major axis' position angle, which perhaps explains the near unity image plane velocity field axis ratio. This would imply a source-plane axis ratio of $\gtrsim 1 / (6 \pm1.5)$ with a major axis position angle of $\sim0^\circ$ east of north, assuming all the magnification occurs along the east-west axis. An attempt was made to include an ellipsoidal source-plane morphology in the source place inversion, however the limited S/N did not allow satisfactory convergence of the MCMC chains. Deeper observations at higher angular resolution will place stronger constraints on the source-plane molecular gas structure.

\subsection{Magnification}

\begin{figure}
\includegraphics[width=0.47\textwidth]{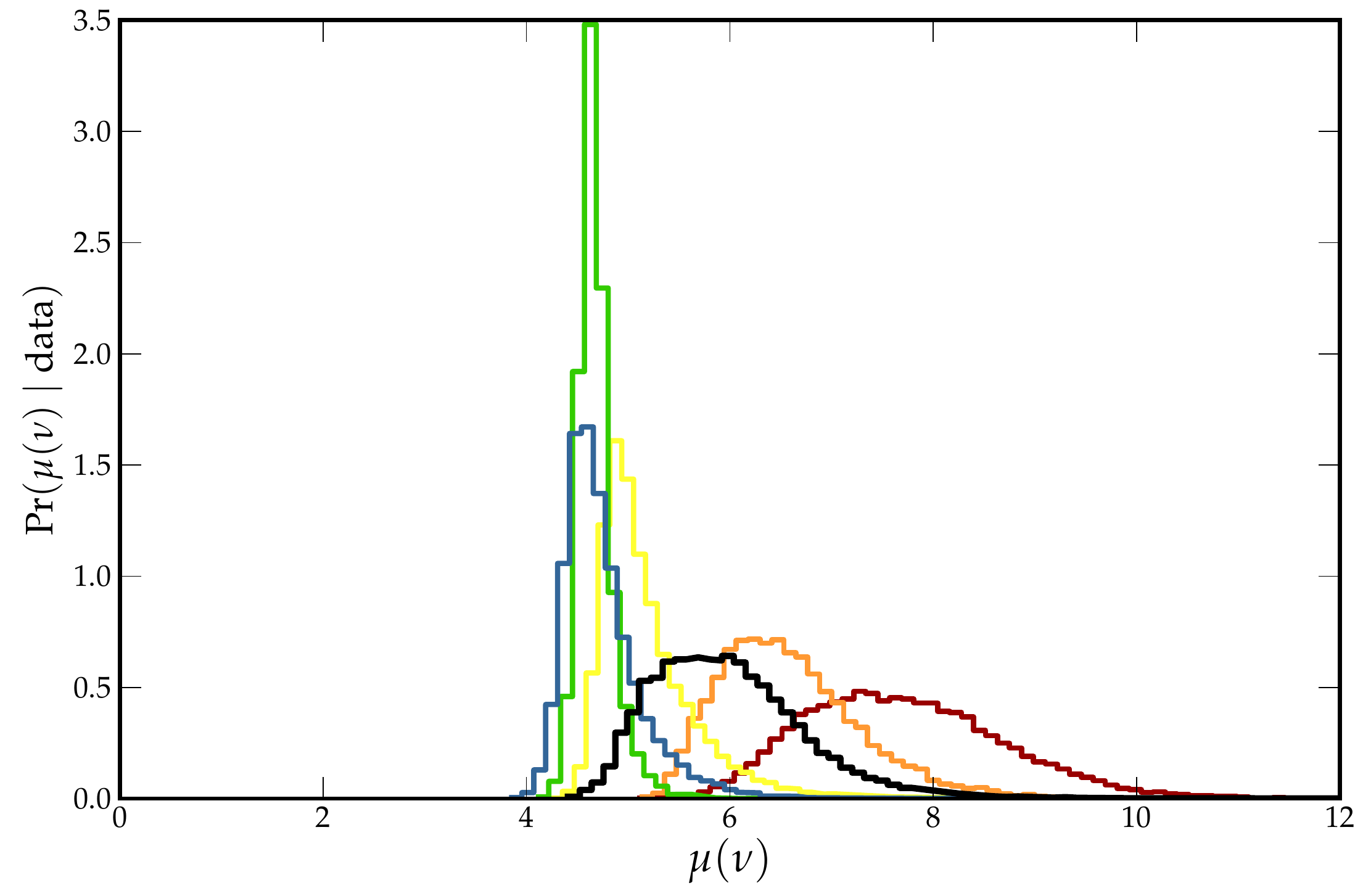}
\caption{Posterior PDFs of individual channel map magnifications (colours). The posterior PDFs are colour-coded by recession velocity, starting with the most blue-shifted channel ($V = -117$ km\,s$^{-1}$) and increasing in steps of $\Delta V = 42$ km\,s$^{-1}$. The black curve represents the total intensity map magnification posterior PDF.    }
\label{fig:munuHist}
\end{figure}

The magnification posterior PDF is sampled by computing the total magnification for each iteration of the MCMC chain. This is calculated by taking the ratio of image to source-plane flux. The procedure is repeated for each channel map with an integrated spectrum flux greater than 5-$\sigma$ ($\sim$1 mJy\,beam$^{-1}$), as well as the total intensity map. In Fig.~\ref{fig:munuHist} we show the MCMC-derived magnification posterior PDFs of each of the 5 channels and the total intensity map. The latter has a mean magnification of $\mu_{\rm CO(1-0)} = 6 \pm 1.5$, which is very similar to the total magnification found for the molecular Einstein ring PSS~{\sl J}2322+1944 ($\mu = 5.34 \pm 0.34$), where the lensing galaxy has a slightly lower Einstein radius ($\theta_{\rm E} = 0.745 \pm 0.024$~arcsec) than in the case of IRAS~10214's main lens ($\theta_{\rm E} = 0.827 \pm 0.044$~arcsec). Given the similar apparent \co luminosity and Einstein ring radii between IRAS~10214 and {\sl J}\,2322+1944, we will compare the properties with the latter (as reported in \citealt{Riechers2008}) throughout this paper. 

The level of preferential magnification shown here is enough to distort the integrated \co spectrum and create the observed asymmetry, however the uncertainties do not provide conclusive evidence for this. \citet{Ao2008} suggest this asymmetry is caused by differential opacity effects for the low vs. high-{\sl J} lines due to the presence of cold foreground material. We have shown in Fig.~\ref{fig:srcPlane} that the red-ward channel centroids are closer to the cusp, and therefore undergo a larger magnification boost. This will contribute towards the observed asymmetry, however a differential opacity function is plausible given the low ($i \gtrsim 30^\circ$) expected inclination of the galaxy.

\subsection{Gas Mass}\label{sec:gasmass}

With the magnification estimated we can now calculate the intrinsic properties of IRAS 10214. We calculate a lensing-corrected \co luminosity of $L'_{\rm CO} = 1.5 \pm 0.3 \times 10^{10}$~K~km\,s$^{-1}$~pc$^2$ with our derived average magnification $\mu_{CO(1-0)} = 6\pm 1.5$. The CO line luminosity $L'_{\rm CO}$ is directly proportional to the total gas mass through

\bea
\centering
M_{\rm gas}({\rm M}_{\odot}) & = & \alpha \, L'_{\rm CO},
\eea

\noindent where $L'_{\rm CO}$ has units K~km\,s$^{-1}$~pc$^2$. The conversion factor $\alpha$ has been determined through dynamical mass measurements in molecular cloud complexes within our Galaxy and external galaxies. The value typically ranges between $\alpha \sim 0.8 - 4.6$~M$_{\odot}$~(K~km\,s$^{-1}$~pc$^2$)$^{-1}$ depending on environment \citep{Downes1998}. For the remainder of the paper, we omit the units of $\alpha$ for brevity. Measurements in the Galaxy are typically close to the upper bound, while a value of $\alpha \sim 0.8$ is found in ULIRGs. We adopt the latter given the infrared luminosity of IRAS~10214 ($L_{\rm FIR} \sim 10^{13} $~L$_{\odot}$). This results in a gas mass $M_{\rm gas} = 1.2 \pm 0.2 \times 10^{10}$~M$_{\odot}$ and implies a 65:1 gas-to-dust ratio based on the dust mass $M_{\rm dust} = 1.1 \times 10^9 \, \mu^{-1} \, $~M$_{\odot}$ derived by \citet{Ao2008} and assuming equal magnifications for the both components. This is consistent with gas-to-dust mass ratios in SMGs \citep[e.g.][]{Kovacs2006}. As we discuss later, the dust component is likely to be more compact than the total gas reservoir \citep{Ao2008}. As a result, emission from dust will undergo a marginally larger magnification according to our lens model. We calculate a gas projected surface density of $\Sigma_{\rm gas} = 408 \pm 80$~M$_{\odot}$\,pc$^{-2}$, based on $>5\sigma$ pixels in the image plane integrated intensity map (note that this calculation is independent of lensing).

\subsection{Dynamical Mass}\label{sec:mdyn}

With spatially-resolved kinematics that are consistent with regularly rotating gas, we can make a tentative estimate of the dynamical mass. The virial theorem states that the dynamical mass $M_{\rm dyn}~=~V^2 \, r / ({\rm G} \, \sin^2 \,i)$, where G is the gravitational constant, $i$ is the disc inclination, $V^2$ is the gas velocity and $r$ is the radius at which that velocity is measured. Caution must be stressed in this section however, since the uncertainties are large and are built on a number of assumptions. We discuss these and derive a dynamical mass inside the effective radius, $M_{\rm dyn}(r < r_{\rm eff}$), based on two methods.

\subsubsection{Integrated Spectrum V$_{\rm FWHM}$ Method}\label{sec:vfwhm}

This method is based on the posterior PDFs of the measured $V_{\rm FWHM}$ of the integrated \co spectrum; and the derived \co effective radius based on our source plane inversion of the \co total intensity map. We follow \citet{Daddi2010} and apply a simulation-derived correction factor that increases the dynamical mass $M_{\rm dyn}$ by 30~percent. This factor results from a number of both negative and positive effects including flattening of the baryonic disk, projection angle bias, non-circular gas orbits (see \citealt{Daddi2010} for more detail). Like these authors, we calculate dynamical mass $M_{\rm dyn}$ inside the effective radius $r_{\rm eff}$ ($\sim 0.7 \, r_{\rm s}$), 

\bea
\label{equ:daddidynmass}
M_{\rm dyn}(r_{\rm eff}) & = & 1.3 \, \frac{ (V_{\rm FWHM}/2)^2 \, r_{\rm eff}}{\rm G \,\sin^2 i}.
\eea

We set the disk inclination $i=90^\circ$ for the moment, but consider a range of inclinations later. We calculate the dynamical mass, $\log(M_{\rm dyn}(r<r_{\rm e})/{\rm M}_{\odot}) = 10.06 \, \pm \, 0.19$ by combining the $V_{\rm FWHM}$ and $r_{\rm eff}$ posterior PDFs (i.e. $M_{\rm dyn}$ is calculated for each sample in the respective MCMC chains). Although the statistical uncertainty above is large, the systematic uncertainty is likely to dominate the error budget. We have assumed that the gas is arranged in a disk and velocities are representative of circular rotation. Furthermore, our inferred radial posterior PDF and the associated magnification (and hence radius) include significant systematics from assumptions such as a singular isothermal ellipsoid potential and the co-location of stellar and dark matter potential centroids (see {\bf D13a} for a full description). Note that this dynamical mass is approximately equal (5 percent less) than the gas mass derived in \S\ref{sec:gasmass}.

\subsubsection{Position-Velocity Fit Method}\label{sec:arctan}

Our second method follows the typical technique used in optical analyses of kinematics in intermediate to high-redshift galaxies \citep[e.g.][]{Miller2011,Swinbank2006}. This collapses the map into a single spatial dimension by assuming the major axis orientation. We determine a best-fit major axis with a straight line fit to the source plane \co channel centroids shown in Fig.~\ref{fig:srcPlane}. These centroids are then used to fit the position-velocity space function,

\bea\label{equ:arctan}
V & = & V_0 + \frac{2}{\pi} \, V_{\rm asyp} \arctan \left( \frac{r - r_0}{r_{\rm t}} \right), 
\eea

\noindent where $V_0$ is the systemic velocity, $V_{\rm asyp}$ is the asymptotic velocity, $r_0$ is the dynamical centre, $r_{\rm t}$ is the turn-over radius (where the second derivative of $V(r)$ is zero). Again, we set disk inclination $i = 90^\circ$ for the moment and return to this at the end of this section. We explore the allowable parameter space with an MCMC algorithm to robustly quantify the uncertainty. We assign uniform priors to all parameters, apart from the systemic velocity for which we employ a Gaussian prior centered on $V_{\rm sys} = -14 \pm 13$~km\,s$^{-1}$, which is based on the integrated spectrum fit.

\begin{figure}
\includegraphics[width=0.47\textwidth]{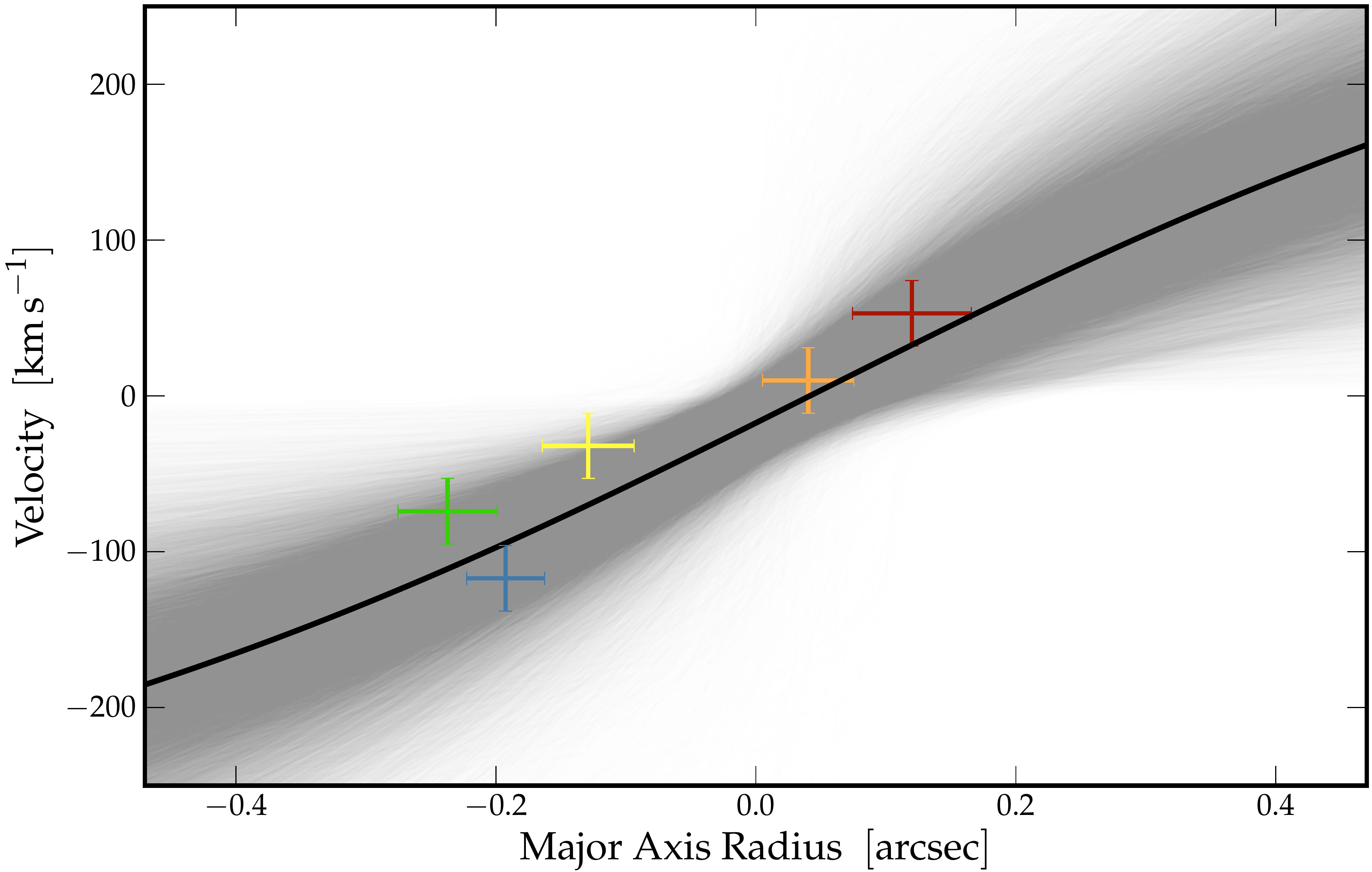} 
\caption{Position-velocity plot of the five channels inverted into the source plane. The data points are coloured by velocity as before. The major axis is derived from a straight line fit to the source-plane channel centroids plotted in Fig.~\ref{fig:srcPlane} and the origin is a linear interpolation to zero velocity ($z=2.2856$). These data are fit to the {\sl arctan} function in Eq.~\ref{equ:arctan}. Every 50$^{\rm th}$ model is over-plotted with low opacity so density is a proxy for probability and so the range of models explored in the MCMC process is also captured in the plot. Note that both the dynamical centre and central velocity are varied in this MCMC algorithm, (i.e. no models with counter-rotation are implied by the figure). The velocity uncertainties represent the channel widths ($\Delta V = 42$~km\,s$^{-1}$) while the major axis uncertainties are MCMC-derived from the lensing-inversion algorithm, and rotated through the major-axis position angle. The red line corresponds to the model with the mean posterior PDF of each parameter and is virtually unchanged for $V_{\rm asymp} \gtrsim 200$~km\,s$^{-1}$. See further details in text.     }
\label{fig:arctan}
\end{figure}

In Fig.~\ref{fig:arctan} we show the position-velocity plot with the five source plane inverted \co channels which are colour-coded by velocity. The uncertainties of these five points reflect the channel velocity width ($\Delta V = 42$~km\,s$^{-1}$); and the MCMC-derived positional uncertainties rotated through the fitted major axis angle. Since the systemic velocity is a free parameter in the fit, we define the origin of the x-ordinate as the linearly interpolated position where velocity (relative to $z = 2.2856$) is zero. Therefore, there are no models with counter-rotation implied by Fig~\ref{fig:arctan}, it simply illustrates the range of models explored during the MCMC fitting procedure. The black line is an {\sl arctan} model derived from the mean of each parameter's posterior PDF. As can be seen in Fig.~\ref{fig:arctan}, we cannot make any constraints on the asymptotic velocity above $V \sim 200$~km\,s$^{-1}$, reflected in an almost uniform posterior PDF. However, some useful constraints can be made on the dynamical mass \emph{within the effective radius}, where $r_{\rm eff} = 4.0$~kpc or 0.48 arcsec, which defines the x-ordinate limits in Fig.~\ref{fig:arctan}. 

We also have poor constraints on the turn-over radius but find a value $r_{\rm t} = 0.6 \pm 0.2$~arcsec (5.0$\pm$1.7~kpc). Comparison with a sample of $z = 0.2 - 1.7$ star forming disk galaxies with similar stellar masses ($10^{10}$\,M$_\odot$; \citealt{Miller2011,Miller2012}) shows that it lies roughly in the centre of the distribution their derived turn-over radii (S.H.~Miller, private comm.). This indicates that it is a plausible value given this scenario, and that the algorithm is returning sensible results despite very few data points.  

\begin{figure}
\includegraphics[width=0.47\textwidth]{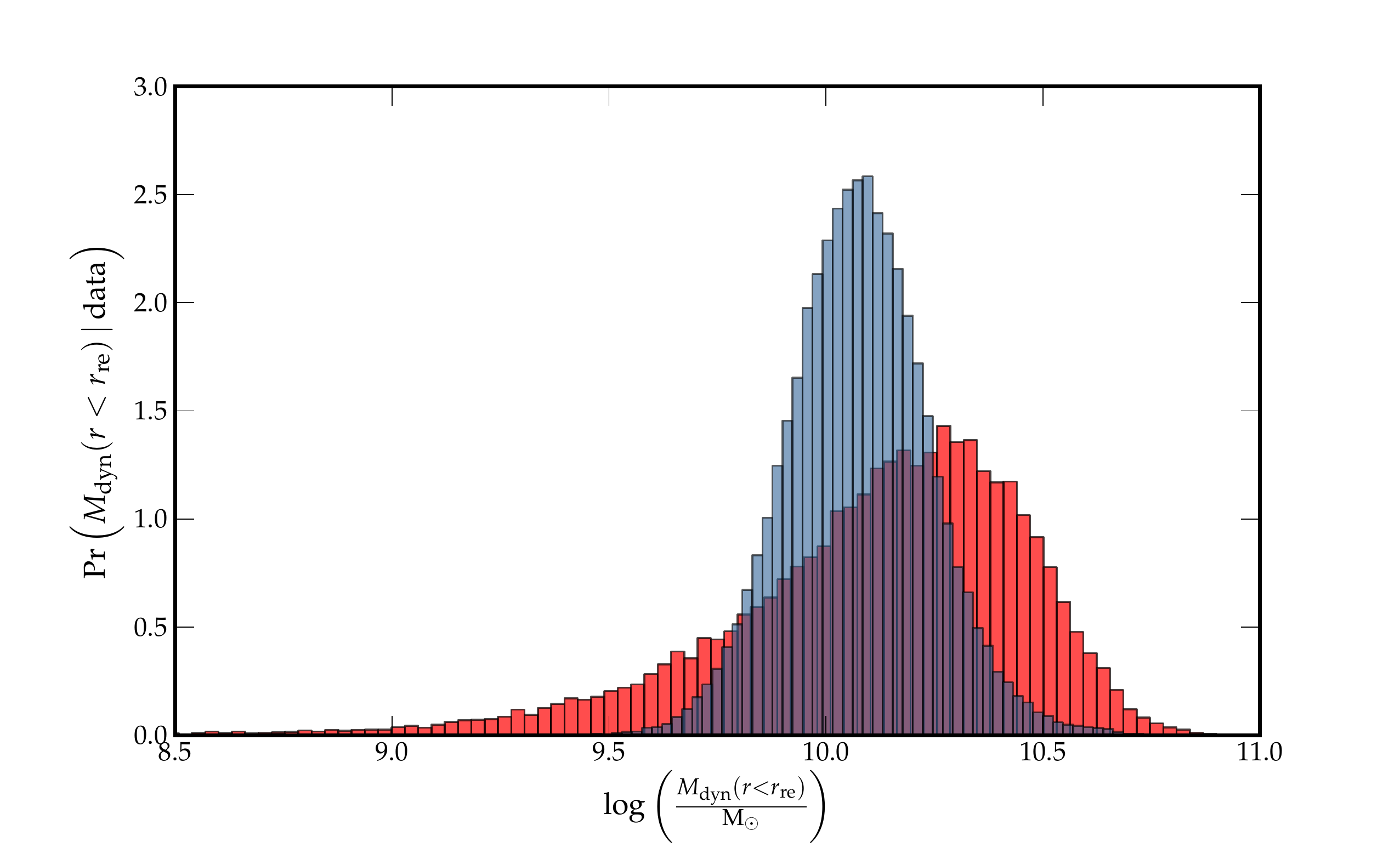}
\caption{Posterior PDFs of the dynamical mass $M_{\rm dyn} (r < r_{\rm e})$ inside the effective radius as estimated by the $V_{\rm FWHM}$ fit to the integrated \co spectrum (blue; see \S\ref{sec:vfwhm}) and the {\sl arctan} model (red; see \S\ref{sec:arctan}).      }
\label{fig:Mdyn}
\end{figure}

Through this MCMC algorithm we are able to sample the $M_{\rm dyn}(r < r_{\rm e})$ posterior PDF by calculating the rotational velocity at the effective radius for each MCMC sample. The result has large uncertainty, however it does provide a consistency check with the Integrated Spectrum $V_{\rm FWHM}$ Method (\S\ref{sec:vfwhm}). In Fig.~\ref{fig:Mdyn} we plot the dynamical mass posterior PDFs of both methods described here. The {\sl arctan} method results in a dynamical mass $\log(M_{\rm dyn}(r<r_{\rm e})/{\rm M}_{\odot}) = 10.28 \,  \pm^{0.16}_{0.48}$. The plot shows these are consistent values, albeit with considerable uncertainty. 

The effect of inclination has not been included in the values quoted above. This increases the dynamical masses by a factor $1/\sin^2(i)$, so here we consider the values for $i = 90^\circ, 60^\circ,$ and $30^\circ$. The latter two inclinations will increase the dynamical mass by factors of 1.3 and 4 respectively. We include these possibilities in our estimation of the $L_{\rm CO}$-to-$M_{\rm gas}$ conversion factor $\alpha$ in the next section. \citet{Lawrence1993} discuss limits on the disk inclination angle based on polarisation arguments which we briefly summarise here. In the case of optically-thin electron scattering, they follow \citet{Brown1977} and \cite{Miller1990} and show that the observed optical/UV polarisation fraction of $>20$~percent requires a disk inclination angle $>40^\circ$. In principle, optically-thin electron scattering can have a polarisation fraction of 100~percent at a 90$^\circ$ inclination angle. However, dust scattering has a maximal polarisation fraction of $\sim35$~percent according to \citet{Scarrott1990}. The high optical polarisation fraction in IRAS~10214 of 28~percent therefore suggests that the disk has a large inclination. Assuming the putative dusty toroidal structure and gas disk are roughly aligned, we can then expect the latter to also be highly inclined. 

The \co FWHM appears to fit well within the scatter of the $L'$({\sc CO (1-0)})-FWHM correlation from 32 SMGs presented in \citet{Bothwell2012}. This, in combination with the appearance that IRAS 10214 has a relatively low gas mass (assuming $\alpha = 0.8$) when compared to the same SMGs, suggests that IRAS 10214 is not unusually narrow, but rather just towards the lower gas and dynamical mass range of FIR luminous galaxies at $z\sim2$. Again, the deep HST {\sl H}-band map (and other \hst filter maps) do not suggest any ongoing major merger, so we would not expect the very wide FWHM associated with such systems (i.e. $> 500$ km\,s$^{-1}$). This is consistent with the finding that the lower FIR luminosity SMGs ($< 10^{13}$~L$_{\odot}$) have much lower merger prevalence \citep[see][]{Targett2012}. We consider inclinations of $30^\circ - 90^\circ$ for the remainder of the paper.

\subsection{Independent  $L_{\rm CO}$-to-$M_{\rm gas}$ Conversion  Estimate}\label{sec:alphaest}

Having approximated the dynamical mass in \S\ref{sec:mdyn}, we can estimate the likely range of the $L_{\rm CO}$-to-$M_{\rm gas}$ conversion factor ($\alpha$) in IRAS~10214. This is of interest since, as previously stated, this value varies by a factor $\sim 6$ depending on environment: $\alpha = 4.6$ in our Galaxy, while the conversion factor has been measured to be $\alpha = 0.8$ in ULIRGs. The latter is used in our gas mass estimate in \S\ref{sec:gasmass}, however if we constrain a number of parameters, we can make an independent measure of $\alpha$. To do so, we assume $M_{\rm dyn} / \sin^2{(i)} = M_{\rm gas} + M_{\rm stellar} + M_{\rm DM}$, where $M_{\rm gas}$ is the total gas mass (neutral + molecular); $M_{\rm stellar}$ is the total stellar mass; $M_{\rm DM}$ is the total dark matter mass; and $M_{\rm dyn}$ is the mean dynamical mass estimated in \S\ref{sec:mdyn} ($\log(\bar{M}_{\rm dyn}) = 10.2 \pm 0.2$). Note that all of the masses in this expression pertain to the radius within which the dynamical mass was estimated, namely $r_{\rm eff,CO} \lesssim 4$~kpc.  
 
We assume a dark matter fraction $f_{\rm DM} =  0.25$ within the \co effective radius \citep[][and references therein]{Daddi2010}. This sub-dominance of dark matter within the effective radius of galaxies, particularly massive disk galaxies, is well-supported by galaxy rotation curves and strong-lensing results \citep{Rubin1980,Treu2004}. An estimate of the stellar mass is more challenging, largely due to the lack of a reliable magnification of the evolved stellar component. \citet{Lacy1998} showed tentative evidence for a Balmer break at 4000 \AA\ in their near-infrared spectrum, suggesting that the host galaxy stellar light may be significant at rest-frame {\sl B-}band, however it seems clear that the near-ultraviolet is dominated by scattered quasar light based on the strong polarisation and the effect of preferential lensing. Ideally, we would make a mass estimate in the near-infrared, however the lens galaxy makes a substantial, yet unconstrained, contribution at this wavelength (Verma et al., in preparation). The complexities require careful, multi-component modelling beyond the scope of this paper. Therefore, we adopt the \citet{Bothwell2012} average baryonic gas fraction ($f_{\rm gas} = M_{\rm gas}/( M_{\rm gas} +  M_{\rm stellar})$), measured in their sample of 32 SMGs with a median redshift of $\langle z \rangle = 2.2$. Their sample has a median gas fraction of $\langle f_{\rm gas} \rangle = 0.43 \pm 0.05$. Here $M_{\rm gas}$ is entirely molecular and assumes that the neutral hydrogen mass ($M_{\rm HI}$) can be neglected within the \co effective radius \citep{Obreschkow2009}. This sample of galaxies is the most appropriate available in the literature, given the size of the sample, the median redshift, median far-infrared luminosity, and the fact that the inferred molecular mass lies within range reported in the \citet{Bothwell2012} sample. However, even if the uncertainty associated with this gas fraction were to significantly increase, the $\alpha$ estimate uncertainty would still be dominated by that associated with the inclination. 

Following these assumptions, the gas mass can be related to the dynamical mass by $M_{\rm gas} = f_{\rm gas} \, (1 - f_{\rm DM}) \, M_{\rm dyn} / \sin^2{(i)}$ = 0.43 (1 - 0.25) 1.58\,$\times 10^{10}/ \sin^2{(i)}  \, {\mathrm M}_\odot  =  5~\times~10^{9} / \sin^2{(i)}$~M$_{\odot}$, within the \co effective radius. This suggests that the stellar mass $M_{\rm stellar} = 6.8 \pm 1.7 \times 10^{9}  \, {\mathrm M}_\odot$, where the uncertainties are the quadrature sum of the gas fraction and \co total flux density uncertainties, however the true uncertainty in $M_{\rm stellar}$ is dominated by the assumptions. The inferred range of $\alpha$ values is therefore $\alpha =   M_{\rm gas}/L'_{CO} = M_{\rm gas,kin}~=~5-20~\times~10^{9}$~M$_{\odot} / 1.5 \times 10^{10}~{\rm K}\,{\rm km}\,{\rm s}^{-1}$ within the effective radius. This implies a range of $\alpha = 0.3 - 1.3$ for inclinations of $i = 90^\circ - 30^\circ$. This range is consistent with the empirically derived value for ULIRGs, providing evidence that this is an appropriate conversion factor for IRAS~10214.

\subsection{Evidence for an Extended Gas Reservoir?}\label{sec:reservoir}

As discussed earlier, observations of molecular gas in high-redshift galaxies have primarily targeted the mid- to high-{\sl J}\, CO lines due to instrumentation and atmospheric window limits. At low redshift, comparisons between low and high-{\sl J} lines have been limited for similar reasons, with high-{\sl J} lines being the more challenging to observe in the nearby Universe (until the launch of the {\sl Herschel Space Observatory}). As a result, definitive comparisons have been difficult to perform which has raised concern about accurate gas mass estimation from the high-{\sl J} line velocity-integrated intensities. This is rooted in the plausible scenario where two gas phases dominate AGN host galaxies with high star formation rates: (1) a higher kinetic temperature phase in the molecular cores and circumnuclear environment where the most intense star formation occurs and/or AGN heating takes place; (2) a lower kinetic temperature gas in the more quiescent parts of the molecular disk which is not directly involved in the central starburst. Since the high-{\sl J} lines would not be very sensitive to the latter component, there could exist a large, subdominant reservoir of low-excitation molecular gas if the lowest {\sl J} line luminosities are not measured. This was particularly true of IRAS~10214 since \coiii was the lowest rotational line detected, and given the higher dust and gas temperatures in this system \citep[e.g.][]{Ao2008}, suggesting that the mid-{\sl J} emission line regions may be predominantly heated by the AGN and intense star formation. 

The \co detection in \citet{Riechers2011} demonstrated that there was no \emph{prima facie} evidence for a large gas reservoir in IRAS~10214, assuming a single temperature component and equal magnification factors for low and high-{\sl J} emission regions. Our observations here spatially resolve the \co line, providing a direct estimate of the apparent solid angle ($\Omega_{\rm app}$) which we will compare with the \coiii solid angle size. In this section, we wish to ask the question: is there evidence for an extended halo of low excitation molecular gas? If so, can we quantify the potential total mass in this extended reservoir? 

To address this, we compare two line excitation models: Model A assumes all CO emission line regions are co-spatial; while Model B uses spatial constraints from the \co map presented here and the \coiii map \citep{Downes1995} to attempt to disentangle the putative high and low excitation emission line regions that are naively expected in a AGN host galaxy with significant star formation. This allows direct comparison with the best-fit solid angle derived from line excitation modelling and constrains the viable range of molecular gas filling factors. We then discuss whether these results are consistent with the \ci\ properties, as well as the multi-component mid-infrared SED modelling.

\subsubsection{Spatial constraints on CO emission regions}\label{sec:spatial}

We do not have sufficient constraints to fit a two-component temperature model to the CO SLED, however, the significant difference in spatial extent of the \co and \coiii maps suggests that we may loosely separate the CO emitting regions into two components: (1) a central, high-excitation emitting region defined by the \coiii extent and loosely referred to as the CO `core'; and (2) an extended, lower excitation emission region defined by the \co extent. 


Using the {\sc aips} task {\sc jmfit}, we determine the deconvolved, image plane major and minor axes $\theta_{\rm maj}, \theta_{\rm min} = 2.7 \times 2.1$~arcsec$^2$ ($r_{\rm e,app} \sim 1.2$~arcsec, $\sim 10$~kpc) for the `extended' component. The approximation of the \co map to a 2D Gaussian is likely to underestimate the complex morphology seen in Fig.~\ref{fig:COmom0}, particularly since the source is dominated by the central component which is co-spatial with the most magnified region in the image plane (see {\bf D13a}).

The Gaussian-fitted solid angle FWHM of the \coiii emission is $\Delta \theta =$1.5$\pm$0.4$ \, \times$$<$$0.9 \ {\rm arcsec}^2$ (apparent spatial FWHM of $D = 12.5 \times < 7.5$~kpc$^2$, \citealt{Downes1995}) and is therefore $< 23$~percent of the deconvolved \co solid angle. Differences in the \co and \coiii solid angles at this level imply significantly different magnification factors ($\Delta \mu \gtrsim 2$) in the case of IRAS~10214, which we return to later. As a point of comparison, the deep \jvla observations of 5 SMGs with very similar median infrared luminosity ($L_{\rm FIR}  \sim 5 \times 10^{12}$\, L$_\odot$) and total gas masses ($M_{\rm gas} \sim 2.5 \times 10^{10}$\, M$_{\odot}$), have much larger \co effective radii ($r_{\rm eff} \sim 8$~kpc; \citealt{Ivison2011}), compared with the \coiii derived effective radii ($r_{\rm eff} \sim 0.8-2.8$~kpc; \citealt{Tacconi2008}).

We use two approaches to calculate the fraction of \co flux that appears co-spatial with the \coiii component. The first assumes the \coiii emission peak is co-spatial with the \co peak; while the second assumes the \coiii peak emission position reported in \citep{Downes1995}. We then apply an aperture of size 1.5$\pm$0.4$\times$$0.9^{+0.0}_{-0.4}$~arcsec$^2$, calculate the \co flux density contained within, and multiply by a factor of 2 since this defines the effective radius for a Gaussian component. This results in a ratio of the \co within the CO `core' to CO `extended' emission components of $f_{\rm core/ex} = 0.46\pm0.09$ and $0.43\pm0.10$ respectively, where the uncertainties are dominated by the \coiii size uncertainty. At face value this implies that roughly half of the \co emission is co-spatial with the \coiii  emission and the other half is located in an extended halo of more diffuse gas. We note that the \co emission co-spatial with the CO `core' is concentrated in an unresolved component and spans an equivalent or even larger range in velocity space the extended emission. This suggests the innermost CO emission component has a broad velocity width and that it is in close proximity to the SMBH, perhaps within the sphere of influence ($r_{\rm SOI} \sim {\rm G}\,M_{\rm BH}/\sigma^2 \sim 100-200$~pc for IRAS~10214), based on the black hole and spheroid mass estimates in {\bf D13c}.

\subsubsection{Line excitation modelling}\label{sec:LEM}

\begin{figure}
\includegraphics[width=0.47\textwidth]{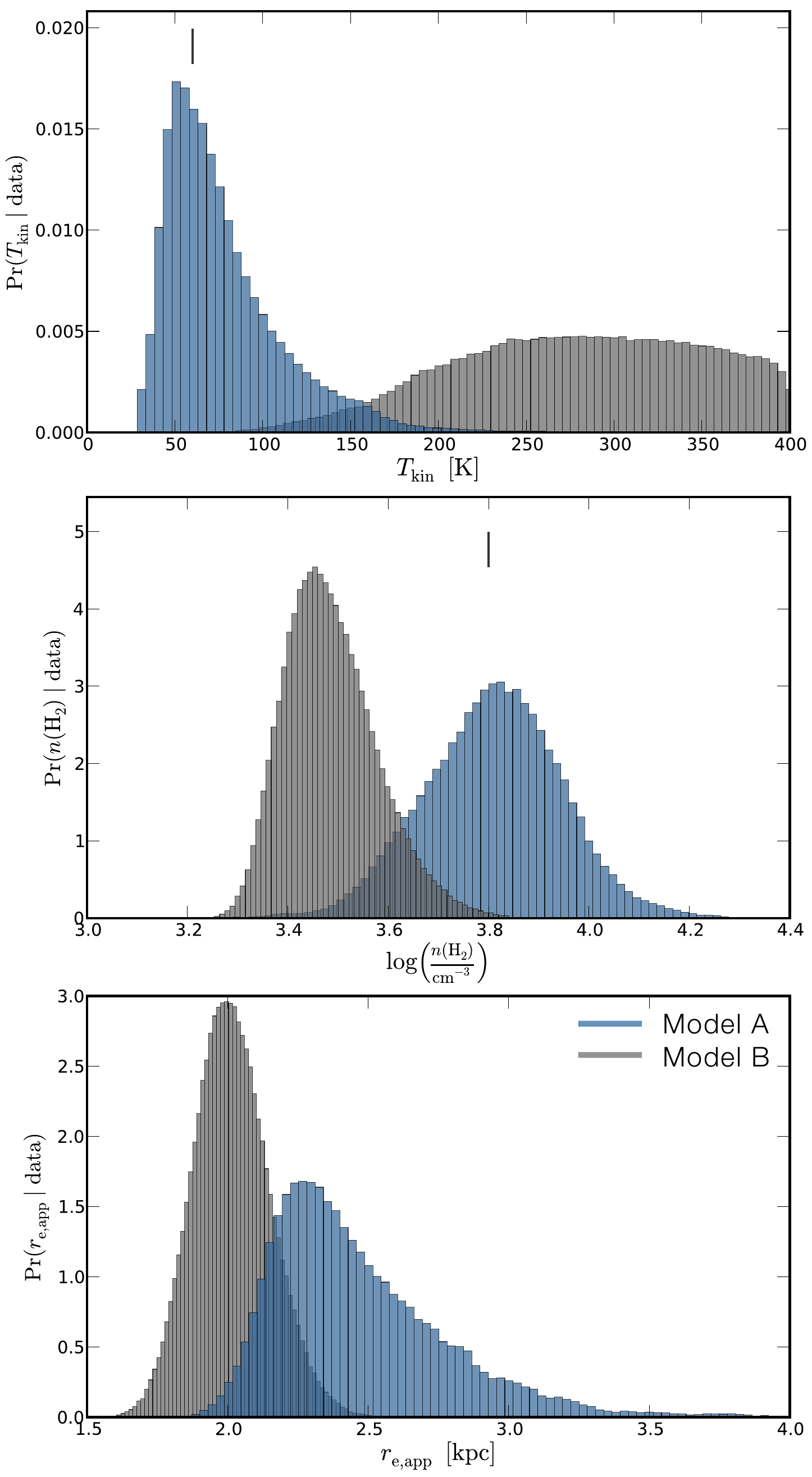} 
\caption{Posterior PDFs of the derived parameters from the LVG modelling for two cases. The blue PDF corresponds to the single temperature LVG model that assumes all \co flux density is co-located with the high-{\sl J} lines. The grey PDF assumes \emph{half} the  \co flux density is co-located with the high-{\sl J} lines, as evidenced by our spatially-resolved \co \evla map. This results in a gas temperature for the high-{\sl J} line gas component of $\gtrsim$250~K, consistent with the dominant hot dust component fitted in the IRAS~10214 IR spectrum \citep{Teplitz2006,Efstathiou2006}. The short black vertical lines indicate the quoted best-fit values for $T_{\rm kin}$ and $n({\rm H}_2)$ in \citet{Ao2008} and \citet{Riechers2011}.  }
\label{fig:LVGpdfs}
\end{figure}

In LVG modelling, the data are the CO flux densities for different rotational quantum numbers, which provide constraints on the average gas properties of individual velocity components (i.e. the line-widths are assumed to be the result of large-scale, systematic motions, beyond plausible thermal velocities within a molecular cloud, \citealt{Goldreich1974}). We use the LVG modelling software developed by Christian Henkel, which is detailed in several papers \citep[e.g.][]{Weiss2007}. We assume a single, spherically symmetric large velocity component; collision rates from \citet{Flower2001} with an ortho-to-para H$_2$ ratio 3:1; and a CO abundance per velocity gradient of [CO]/\mbox{(${\rm d} v / {\rm d}r)$} $= 10^{-5}$~pc\,(km\,s$^{-1}$)$^{-1}$. This software generates a library of line brightness temperatures for a given H$_2$ density ($n({\rm H}_2)$) and gas kinetic temperature ($T_{\rm kin}$). These brightness temperatures are converted to a corresponding CO line intensity (velocity integrated flux density) through the relation

\bea 
S_{\rm CO} & = & \frac{2 k}{c^2} \, T_{\rm b} \, \Omega_{\rm app} \, \nu^2_{\rm obs} \, (1 + z)^{-1}, 
\eea

\noindent where $T_{\rm b}$ is the line brightness temperature, $\Omega_{\rm app}$ is the apparent (i.e. magnified) solid angle of the source and $k$ is Boltzman's constant. \citet{Weiss2007} defined the equivalent radius $r_0$ assuming a uniform, face-on disk ($r_0 = D_{\rm A}\,\sqrt{\Omega/\pi}$). To facilitate comparison with our source-plane and image-plane radii, both of which are assumed to have Gaussian profiles, we use the effective radius. Since the latter is defined as the radius within which half the total light is contained, we perform our LVG modelling by dividing the model intensities by a factor 2. The derived solid angle is then converted into an effective radius by assuming circular symmetry. We can therefore directly compare LVG output radii with the MCMC-derived \co source-plane effective radius $r_{\rm eff}$  as well as the deconvolved image-plane effective radius $r_{\rm eff,app}$. Since we wish to have well-defined uncertainties, we have incorporated the radiative transfer modelling into an MCMC algorithm in order to output the marginalised (and covariant) uncertainty for each (pair of) free parameter(s). 

Based on the \co and \coiii spatial constraints, we perform two tests to investigate if there may be multiple excitation components in IRAS~10214. In Model A we fit the CO SLED with a single temperature (i.e. assume all CO emission arises from the same region which is in clear violation of the spatial constraints) and find $T_{\rm kin}$ and $n({\rm H}_2)$ posterior PDFs that peak at the values quoted in previous work ($T_{\rm kin} \sim 60$~K, $n({\rm H}_2) \sim 10^{3.8}$~cm$^{-2}$; see \citealt{Ao2008,Riechers2011}). This results in an apparent effective radius $r^{\rm LVG}_{\rm eff,app} \sim 2.3$~kpc ($\sim$0.28 arcsec), which is smaller but consistent with the observed \coiii Half Width Half Maximum (HWHM) $r_{\rm 3-2,HWHM} = 0.7 \pm0.4 \times < 0.45$~arcsec, however it is considerably smaller than the extended \co component radius $r_{\rm e,app} \sim 10$~kpc (1.2~arcsec) \co extent quoted earlier. This is roughly a factor $\sim4.5\times$ greater than the LVG-derived effective radius (factor $\sim20 \times$ greater in solid angle). The LVG-predicted apparent effective radius requires a CO filling factor for comparison (indeed, their comparison sets limits on the filling factor). Assuming a near unity filling factor, the observations appear inconsistent with the modelled radius which is not discussed in \citet{Ao2008} and \citet{Riechers2011}.

The molecular gas radius degeneracy with filling factor is discussed in the following section, however, we now explore a second model which is motivated by the significantly different \co and \coiii radii. Model~B is an exact repeat of Model~A, however the \co intensity is not the total spatial and velocity integrated value, but rather the total \co intensity within the \citet{Downes1995} constrained \coiii size as described in \S\ref{sec:spatial}. Therefore, for Model B, we assume that all the CO ($J \geq$ 3) is co-spatial and make the rough approximation that only 50~percent of the \co flux density is within this region. This is clearly a relatively crude experiment, however we wish to explore effects on the LVG modelling results, \emph{given the spatial information} from the \co and \coiii emission and compare these with the global LVG modelling results performed previously by \citet{Ao2008} and \citet{Riechers2011} and repeated here.

\begin{figure}
\includegraphics[width=0.45\textwidth]{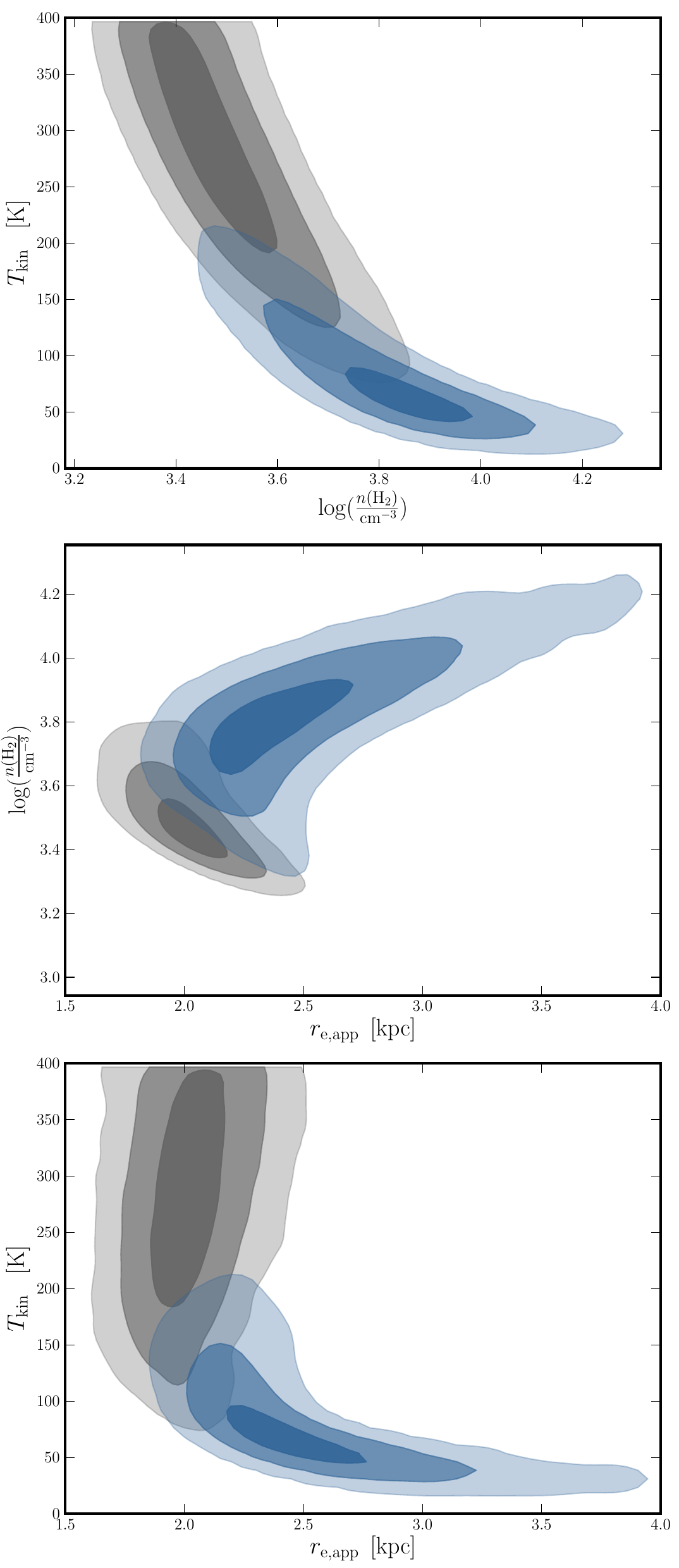}
\caption{Two-dimenisional posterior PDFs of the derived parameters from the LVG modelling for Model~A and B.  (colours as in Fig.~\ref{fig:LVGpdfs}). The shading shows the 68, 95 and 99~percent confidence intervals in decreasing opacity. The top panel shows the commonly shown degeneracy between gas density and kinetic temperature.} 
\label{fig:2DLVGpdfs}
\end{figure}

The resultant posterior PDFs of $T_{\rm kin}, \ n({\rm H_2}), \ $ and $r_{\rm eff,app}$ for both tests are shown in Fig.~\ref{fig:LVGpdfs}. The most dramatic change here is in the inferred kinetic temperature of the single large velocity component. The constraints on this parameter drop dramatically, but now peak at 250-300~K, which is similar to the $\sim200$~K dust component fitted to the IR spectrum. This second dust component has substantial luminosity, comprising of 50~percent of the total IR luminosity \citep{Teplitz2006,Efstathiou2006} which is similar to our (lensed) cold-to-warm gas ratio of 50~percent implied here. Note that the generated LVG library is limited to temperatures $< 400$~K, so the plausible range of kinetic temperatures could be even higher, implying that the posterior probabilities in Model~B are likely to be lower. 

\begin{figure}
\includegraphics[width=0.47\textwidth]{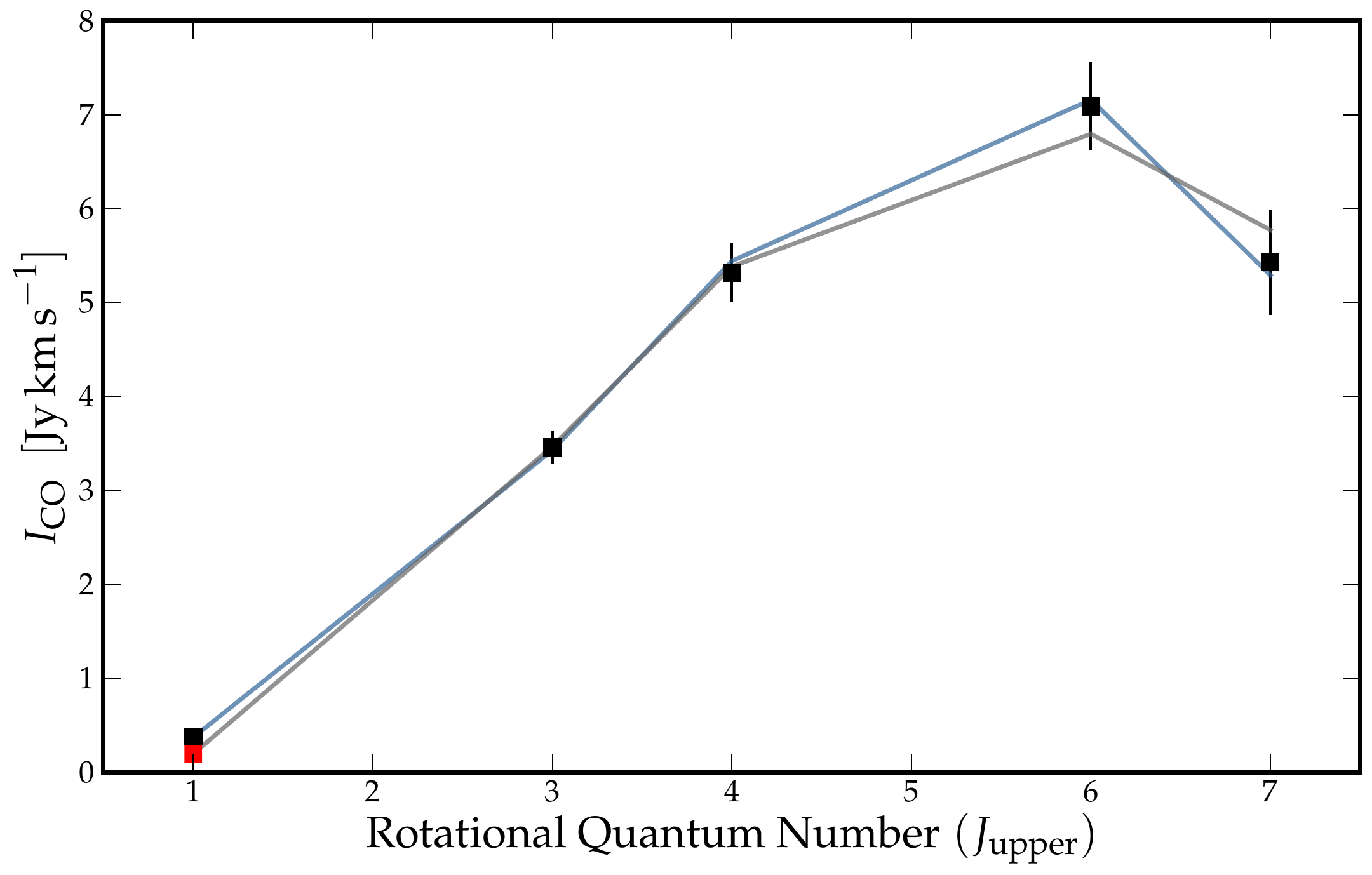}
\caption{ CO SLED fits from LVG Model A and B (blue and grey respectively). The blue curve corresponds to the single temperature LVG model that assumes all \co flux density is co-located with the high-{\sl J} lines. The grey curve assumes \emph{half} the  \co flux density is co-located with the high-{\sl J} lines (see resultant red \co data point), as evidenced by our spatially resolved \co \evla map. This small change demonstrates the importance of accurate low-{\sl J} intensity measurements as shown in Fig.~\ref{fig:LVGpdfs}.  }
\label{fig:CO_SLED}
\end{figure}

In Fig.~\ref{fig:2DLVGpdfs} we show the two-dimensional posterior PDFs of  $T_{\rm kin}, \ n({\rm H_2}), \ $ and $r_{\rm eff,app}$ for both Model~A and Model~B results. In Fig.~\ref{fig:CO_SLED} we plot the two CO spectral line model fits, with the data points taken from the higher-{\sl J} observations from \citet{Ao2008} and the combined \co intensity from this work as well as the \citet{Riechers2011} unresolved \gbt and \evla spectra. The resultant fit of the two tests can be compared in two ways: the minimum reduced-$\chi^2$ or the MCMC chain average reduced-$\chi^2$, where the $\chi^2$ is generated from the residuals of the data-model comparison for all observed CO rotational lines.

Model A results in a minimum reduced- $\chi^2$ that is 50 percent of that achieved in Model B, while both values lie significantly below 1 (0.1 and 0.2 respectively), suggesting that the data are being over-fitted somewhat. However, if the average reduced- $\chi^2$ is considered, then Model B has a 20 percent lower value (1.4) than Model~A (1.7). The related mean log likelihood is -0.85 and -0.7 for Models A and B respectively. As shown by \citet{Lewis2002}, this quantity serves as a proxy for the Bayesian evidence, suggesting that the data are only very marginally more probable given Model B. The point here is simply to demonstrate that the two models can fit the data similarly well, despite the modification made to the CO SLED based on the spatial information. Ruling out one of these models would require higher S/N observations, including more line measurements along the CO ladder (particularly at the high end). Nonetheless, the evidence for an extended reservoir comes from the spatial information and \emph{not} the quality of the LVG fits.

\subsubsection{Degeneracy with molecular gas filling factor}\label{sec:fillingfactor}

The discrepancy between the measured and LVG-derived molecular gas radii could be reconciled with CO filling factors of $f_{\rm CO1-0} = 0.05$ and $f_{\rm CO3-2} = 0.22$, as calculated from the ratio of the Model~A predicted radius to the CO emission solid angles listed in \ref{sec:spatial}. While the filling factor is difficult to constrain due to limited angular resolution and brightness temperature sensitivity, we discuss some results of the literature that are relevant to the case of IRAS~10214.

\citet{Downes1993} argue that volume filling factors cannot be as low as 0.001-0.01 since the resultant true gas densities would be implausible given the observed CO line ratios in (U)LIRGs. They suggest a volume filling factor of 0.3-0.7 for typical ULIRGs, which would only increase if there was any kind of flattened and inclined morphology. This range is consistent with the filling factor of 0.5 derived for the diffuse component in a two phase model of a $z \sim 4$ SMG \citep{Carilli2010}. A high filling factor is determined for a $z\sim 5.3$ SMG detected by \citet{Riechers2010} who find $f  \gtrsim 0.75$ and $f  \gtrsim 0.1$ for the low and high excitation components respectively. 
Furthermore, we can constrain the range of filling factor values based on line excitation modelling results of a number of $z \sim 3$ SMGs of comparable molecular mass to IRAS~10214, where the relative filling factor between the low and high excitation molecular gas components was estimated following multiple component modelling of their CO SLEDs \citep{Riechers2011SMGs}. These authors find that the low excitation components have significantly higher surface filling factors relative to the higher excitation components by factors of 8-20. This is contrary to what is required in IRAS~10214 in order to reconcile the \co and LVG-derived molecular gas radii. More robust constraints on the molecular gas filling factors in IRAS~10214 will require higher resolution, high brightness temperature sensitivity observations centred on multiple CO lines.



\subsubsection{Consistency with \ci\ and mid-infrared SED models}

Given that the \co line luminosity is directly proportional to the gas mass, we can argue that roughly half the total gas mass is in the extended ($r_{\rm eff,app} \sim 10$~kpc) emission detected in the \co total intensity map (assuming a constant $\alpha$). The fraction of mass in this extended component could be even larger, since it appears to undergo a lower magnification boost (by a factor $\sim$2) than the compact CO `core' (see Fig.~4 in {\bf D13c}). In addition, the conversion factor $\alpha$, of luminosity to gas mass ratio could conceivably be larger in the more extended, quiescent extended environment as discussed in \citet{Carilli2010}, given the dependence on metallicity and pressure \citep[e.g.][]{Genzel2012,Blitz2006}.  

A more quiescent, cold gas reservoir is consistent with the derived \ci\ gas excitation temperature $T_{\rm ex,CI}~=~45$~K, which is expected to be co-spatial with CO\,{\sc (1$\rightarrow$0)}, as has been been demonstrated in the Galactic centre and a number of nearby galaxies \citep[e.g.][]{Schneider2003,Israel2002}. The co-location of \ci\ and \co in IRAS~10214 is supported by their similar line-widths, however this should be interpreted with caution given that \ci\ is optically thin. The excitation temperature inferred from the ratio of the two \ci\ hyper-fine transition line luminosities is free of lensing distortions and so provides a robust estimate of the quiescent gas temperature. This is significantly lower the CO SLED fit of $T_{\rm kin,CO} \sim 60$~K and the dust temperature which is $T_{\rm dust} = 80 \pm 10$~K when fitted with a single grey-body component between 60-3000~$\mu$m \citep{Ao2008}.

So the overall picture suggests that unless the filling factor is $f_{\rm CO(1-0)} \lesssim 0.05$, there may be an extended reservoir not captured by modelling the CO SLED when just a single gas phase is considered. This decoupling of the gas into a preferentially magnified `high-{\sl J}' region and an extended disk of more quiescent gas is consistent with the multi-component infrared models; the \ci\ excitation temperature; the \citet{Downes1995} constrained \coiii size; and more generally, the recent \evla observations of submillimetre galaxies (SMGs). 

As discussed in \citet{Obreschkow2009heuristic}, the ability of LVG models to successfully reproduce CO SLEDs (within the usual relatively large uncertainties) can largely be attributed to the fact that most CO SLEDs are dominated by a single gas phase. Indeed, the example of M\,82 reveals that different gas phases dominate in different regions in the galaxy once sufficiently high S/N and spatial resolution is achieved \citep{Weiss2005,Panuzzo2010}. Since IRAS~10214 shows both AGN and starburst traits at most observing frequencies it is not unexpected that there may be more than a single component present. Based on the \co and \coiii size estimates for IRAS~10214 (and for similar SMGs at $z \sim 2$), it appears that the preferentially lensed AGN region gives the higher excitation gas phase a magnification boost of order 2-3$\times$ over the lower excitation gas (see Fig.~4, {\bf D13c}). Hence, we would likely fit a lower single phase gas temperature if the CO SLED were correctly de-lensed, which is again consistent with the lensing independent \ci\ excitation temperature of $T_{\rm ex,CI} = 45$~K, however, we note that the two emission components need not be in thermal equilibrium.

\subsubsection{Speculative implications}

These results show the importance of accurate low-{\sl J}~line fluxes in line excitation modelling. They also suggest how preferential lensing of one part of the CO SLED can dramatically change the inferred properties. This raises some concern given that most LVG modelling of high-redshift sources has been performed on lensed objects and highlights how accurate lens models are essential to tackle these large systematic uncertainties. A starting point to separate the effect of preferential lensing and its distortion to CO SLEDs is provided in \citet[][Sec.~3]{Obreschkow2009heuristic}. These authors apply prescriptions of AGN and starburst heating to the excitation temperature based on semi-empirical results. This type of framework would allow the results of lensing analysis and resolved spatial information to be incorporated into LVG models.

This paper is therefore a first attempt to reconcile the predicted and observed extent of the cold molecular gas in IRAS 10214, given the first reasonably resolved \co map of the system. Although the models are reasonably fit with single temperature components, much higher S/N observations with more extensive coverage of the CO ladder are beginning to show a number of cases where multiple components are necessary. These results appear to be consistent with the mounting evidence that many SMGs have lower excitation, extended gas components that increase the mass estimates by factors of ~2-3 \citep[e.g.][]{Ivison2011, Carilli2010, Danielson2011, Harris2010, Riechers2011SMGs}. A local example of this is given in \citet{Panuzzo2010} who modeled the CO SLED of M\,82 based on {\sl J} lines from J$_{\rm upper}$ = 1 to 13. 


So while single temperature LVG modelling will always work well if the gas is dominated by a characteristic phase, sub-dominant, low excitation phases (not associated with the central starburst) may only be revealed with spatial information. Analogous to the preferentially magnified AGN masking the expected PAH emission in IRAS~10214, we speculate that the preferentially magnified high-{\sl J} CO lines mask the lower excitation component observed in similar non-lensed galaxies as well as starbursts in the local Universe. However, these assertions can only be irrefutably confirmed with higher angular resolution observations of the molecular gas.

\section{Discussion}

\subsection{General}

This spatially-resolved \co map contributes significantly to the overall understanding of IRAS~10214. The total intensity and velocity field maps are consistent with ordered rotation on a $\sim$5~kpc scale. These two maps simultaneously show evidence for some minor merger activity based on the secondary \co components ($A$ and $B$ in Fig.~\ref{fig:COmom0}) and the higher velocity associated with component $A$ in the velocity field (Fig.~\ref{fig:COmom1}), however no other wavelength has provided direct evidence of merger activity in IRAS~10214 to date. 

A source plane reconstruction of five individual channel maps reveals a relatively linear orientation of their true centroid positions, as would be expected if the velocity field is assumed to trace regularly rotating gas. The previously derived positions of the AGN core and NLR lie in the centre of this \co source plane structure (i.e. co-spatial with the velocity channel centroid nearest to the systemic velocity, see Fig.~\ref{fig:srcPlane}). The overall \co magnification indicates the starburst in this galaxy is significantly less magnified than the AGN and narrow-line region as probed by the \hst and radio maps in {\bf D13a}. This preferential magnification is greater than a factor 3 in the case of the narrow line region where we expect hot dust $T_{\rm dust} \sim 200 - 600$~K to be located. This is the likely explanation of the lack of PAH features in the mid-IR {\sl Spitzer} spectrum, which would be dominated by hot dust continuum emission under our preferential lensing scenario. \citet{Teplitz2006} suggest that preferential magnification of the AGN by a factor $\sim$3$\times$ could suppress these PAH features, in line with the level we determine here. The integrated \co spectrum is asymmetric with a more prominent blue-wing as seen in the \coiii and \ci~ spectra \citep{Downes1995,Ao2008}, but not in the higher-{\sl J} CO lines. This could be as a result the preferentially lensed red-ward channels, however the spatially-integrated channel sensitivity limits the significance of this result.

Despite all of the above uncertainties, our derived estimate of the CO luminosity to gas mass ratio has a mean consistent with the canonical value derived for nearby ULIRGs of $\alpha \sim 0.8$. However, recent observations of \co in SMGs and ULIRGs have shown that the value of $\alpha$ may be higher based on spatial and dynamical mass arguments \citep[e.g.][]{Harris2010,Danielson2011,Ivison2011}. These works suggest values of $\alpha$ that are 2-3$\times$ larger than the canonical value of 0.8. Nonetheless, our derived range of $\alpha$ values, builds a picture of consistency compared with high-redshift ULIRGs as well as a vote of confidence in the source-plane reconstruction.

\subsection{Redshifts}\label{sec:redshifts}

\begin{table*}
\label{tab:redshifts}
\centering
\begin{tabular}{l  l  l l  l l l}
\hline

{\bf Component} & {\bf Tracer} &  {\bf Redshift} & {\bf Uncertainty} &{\bf Velocity$^\dagger$}  &  {\bf $V_{\rm FWHM}$} & {\bf Reference}  \\
                &              &                 &                   &       km\,s$^{-1}$       &    km\,s$^{-1}$      & \\
\hline
\hline

BLR             &    \civ\ (polarised)                &      2.2790    & -         & -603 & 6000   & \citet{Goodrich1996} \\ 
BLR             &    \civ                            &      2.2833     & -        & -210  & 1100   & \citet{Goodrich1996}        \\ 
NLR             &    \oiii$\lambda$5007              &      2.2853   &  0.0001 &   -36   & 1040   & \citet{Lacy1998}     \\ 
NLR (?)          &    {\sl Ly-}$\alpha$ (blue-wing)   &      2.2790  &  0.001 &   -603  & 0-900   & \citet{Lacy1998}  (double Gaussian)   \\ 
                 &    {\sl Ly-}$\alpha$ (red-wing)    &      2.2920  &  0.001 & +581  & 0-600   & \citet{Lacy1998}   (poor fit)  \\ 
Dense Gas            &    HCN (1$\rightarrow$0) &     2.2858          & 0.0002   & +18    & 140$\pm$30 & \citet{VandenBout2004}  \\
Warm Gas             &    CO (7-6)                &      2.28534      & 0.00005  & -24     & 248$\pm$30     & \cite{Ao2008} \\ 
                     &    CO (3-2)                &      2.28535      & 0.00005  & -23     & 199$\pm$13     & \cite{Ao2008} \\ 
                     &    CO (3-2)                &      2.2854      & 0.0001  & -18     & 220$\pm$30     & \cite{Downes1995} \\ 
Cold Gas             &    \co                     &      2.28554      & 0.00005  &  -14       & 194$\pm$31       & this work \\ 
                     &    \co                      &      2.2856      & 0.0001   &   0      &   184$\pm$29          &  \citet{Riechers2011} (\gbt values) \\
                                           &   \cifull                &     2.2854         & 0.0001   & -18            & 182$\pm$30 & \citet{Ao2008}  \\
                      &   \cifullio               &     2.2854        & 0.0001   & -18            & 160$\pm$30 & \citet{Weiss2005}  \\ 

\hline
\end{tabular}
   
   \caption{Redshifts of Physical Components in IRAS 10214. \newline \,$^\dagger$Relative to the systemic velocity assumed for this paper, $z_{\rm sys} = 2.2856$, derived from the average integrated CO spectrum in \citet{Ao2008} which has the lowest uncertainty of all redshift determinations, however does not include Doppler tracking uncertainties.} \label{tab:redshifts}
\end{table*}

In Table~\ref{tab:redshifts} we collate a number of redshifts from the literature. We select the highest S/N detections of lines that can be approximated to trace specific physical components. It appears that the canonical redshift of $z~=~2.2856$ is the most accurate and least biased measurement. This is based on the averaged CO spectrum from \citep{Ao2008}, which is consistent with all the molecular line redshifts and the NLR tracers which have blue-shifts. The \oiii\ line is centered at $V_{c} = -36$~km\,s$^{-1}$, while the blue- and redshifted  {\sl Ly}-$\alpha$ components are symmetric around the molecular gas at velocities of $V_{Ly-\alpha, \rm blue} = -603$~km\,s$^{-1}$ and $V_{Ly-\alpha,\rm red} = +581$~km\,s$^{-1}$ respectively. The two {\sl Ly}-$\alpha$ components have comparable flux (the blue component is fractionally greater), suggesting that these are mirrored components of a resonantly scattered {\sl Ly-}$\alpha$ line centered at $z = 2.2856$. At face value, this is supported by a 1:1 \ovi\ doublet ratio, however, it implies exceptionally large densities that are only found in the broad-line region. The most probable explanation for the unlikely \ovi\ doublet ratio is significant contamination by {\sl Ly-}$\beta$ absorption as discussed in {\bf D13a} and detailed in \citet{Serjeant1998}. If these are indeed the resonantly scattered {\sl Ly-}$\alpha$ wings, we can estimate the hydrogen column density ($N_{\rm H}$) required to create the observed velocity offset of each peak through \citet[][Eq.~1]{Neufeld1988},

\bea
\left( \frac{V_{\rm d}}{  \mathrm{km\,s}^{-1}} \right)  & = & 195 \, \left( \frac{N_{\rm H}}{10^{20} \,  \mathrm{cm}^{-2}} \right)^{\frac{1}{3}} \ \left( \frac{T}{10^4 \, {\rm K}} \right)^{- \frac{1}{6}} 
\eea

\noindent where $V_{\rm d}$ is the observed velocity offset of a {\sl Ly-}$\alpha$ peak, $T$ is the gas temperature, which we assign the average kinetic temperature $T_{\rm kin,CO} = 60$~K, determined by the CO SLED (this work and \citealt{Ao2008}). This results in a hydrogen column density $N_{\rm H} = 4 \times 10^{22}$~cm$^{-2}$. This can be considered a rough lower limit given the weak dependence on gas temperature and considering that the emergent {\sl Ly-}$\alpha$ emission is likely to be anisotropic with greater prominence along the jet axis. As a result, the measured velocity peak position may have an inclination dependence. This does not imply that $N_{\rm H} \propto \sin^3(i)$, however it does suggest that the inferred hydrogen column density should increase if any anisotropy exists.

Inspecting the molecular gas kinematics, it appears that the line-width of the higher-{\sl J} CO\,(7$\rightarrow$6) line is larger than that of the lower-{\sl J} \co and \ci\ lines. The FWHM weighted mean from all CO lines with upper {\sl J} $>$3 is $\langle$FWHM$_{\rm warm}\rangle$ = 220$\pm$11~km\,s$^{-1}$, while the colder gas has a $\langle$FWHM$_{\rm cold}\rangle$ = 178$\pm$14~km\,s$^{-1}$. Naively we might expect the opposite if AGN heating contributed to the higher-{\sl J} line flux densities, however this may be a low S/N artifact. In a system not entirely disrupted by a major merger, we would expect the most dense star-forming regions to be concentrated in the galaxy centre. These two effects would lead to larger line-widths for the more extended, colder gas since our dynamical modelling shows that the \co effective radius is well within the radius at which the rotation velocity flattens. However, this is subject to opacity of the respective CO lines as well as the fact that line-widths are potentially influenced by the more complex conditions found at higher densities and temperatures near the active core and central starburst. This is particularly true if jet outflows are present, evidence of which was presented in {\bf D13a}. In addition, the higher density and temperature gas may partially lie within the SMBH sphere of gravitational influence ($r_{\rm SOI} \sim 100-200$~pc as derived earlier), a region we expect to be preferentially magnified based on our lensing inversion of the VLBI detected radio core ({\bf D13c}). 

In summary, there are a number of physical mechanisms that could lead to differing linewidths for the low and high {\sl J} CO lines, including differing opacities, AGN heating, AGN jets or outflows, preferentially magnification near the AGN, differing radial profiles and clumpiness between low and high-{\sl J} lines. To disentangle these would require high resolution observations of the higher {\sl J} CO lines.

\subsection{IRAS 10214 as an Archetype ULIRG}

The source plane structure shows that if higher-{\sl J\,} CO lines are confined to smaller solid angles around the AGN core (as we would expect since the high temperatures require AGN heating) then the CO line spectral energy distribution will be significantly distorted, with higher-{\sl J} lines undergoing magnifications that are $\gtrsim 2-3 \times$ larger than the \co by virtue of their size and proximity to the caustic. Further indirect evidence for preferentially lensed high-{\sl J} lines and dust are the differences in dust and gas temperatures measured in \citet{Ao2008}, however, we note that the dominant gas phase and dust need not be thermally coupled \citep[e.g.][]{Panuzzo2010}. \citet{Ao2008} find that the 1.3 mm continuum emission region is smaller in size than the atomic and molecular gas, which could suggest some AGN heating contribution assuming the dust and gas are co-spatial, however this could also be the signature of a concentrated circumnuclear starburst.

Since IRAS~10214 is frequently used an archetype ULIRG for comparisons of mid-IR spectra, CO SLEDs and infrared luminosities, we argue this is not a sound choice until the function $\mu (\nu)$\footnote{Although gravitational lensing is achromatic, different emission scales have differing magnification factors that can result in a distortion of the global SED. The function $\mu (\nu)$ can be simplified as the mean magnification of the dominant emission component at a given frequency.} has been defined for each wavelength of interest. This strongly motivates detailed comparison analyses of non-lensed objects \citep[e.g.][]{Schumacher2012} in parallel with the sensitivity-enhanced observations that strongly-lensed systems offer.

\section{Conclusions}

We have performed deep \evla {\sl Ka}-band observations centered on the \co line towards IRAS 10214, a $z=2.3$, gravitationally lensed ULIRG harbouring a hidden quasar. We spatially and spectrally resolve the \co emission in IRAS~10214 and use our previously derived lens model to invert five \co channels into the source plane. This enables a number of lines of enquiry not available to unresolved detections and illustrates the unique ability of radio telescopes to spatially resolve cold gas at high redshift. We make the following conclusions.

\begin{enumerate}

\item Our \co map reveals a clear arc-like structure that is co-spatial, but considerably larger than the \hst-traced emission (in all optical/near-infrared filters). 

\item We measure a lensing corrected \co luminosity of $L'_{\rm CO} = 1.5 \pm 0.3 \times 10^{10}$~K~km\,s$^{-1}$\,pc$^2$ which implies a total gas mass $M_{\rm gas}~=~1.2 \pm 0.2 \times~10^{10}$~M$_{\odot}$, if we assume the standard CO luminosity to gas mass conversion factor for ULIRGs ($\alpha \sim 0.8$). 

\item There is a 3-$\sigma$ \co counter-image detection that is co-spatial with counter-images seen through the same \hst filters. This leads to a \co arc to counter-image flux ratio of $\check{\mu}~\sim~7~\pm^6_2$. 

\item The intensity-weighted mean velocity field reveals some order, suggestive of regularly rotating gas with a FWHM $\sim$200~km\,s$^{-1}$. However, the total intensity map does suggest minor merging activity may be present or complex morphology. This is in contrast to major mergers observed in other high-redshift ULIRGs as evidenced by large CO velocity widths of 500-1000 km\,s$^{-1}$ \citep[e.g.][]{Frayer2000,Greve2005}. 

\item Based on our source plane inversion of 5 channels, we estimate a dynamical mass range within the \co effective radius. Our independent estimate of the CO luminosity to gas mass conversion factor $\alpha$ gives a range $\alpha = 0.3 - 1.3$ for inclinations of $i = 90^\circ - 30^\circ$, straddling the typical ULIRG value of $\alpha = 0.8$.

\item The \co emission is a factor 3 less magnified than NLR component as presented in {\bf D13a}, and factor 11 less magnified than the VLBI radio core presented in {\bf D13c}. Therefore, assuming the higher-{\sl J} CO lines are associated with the AGN and not galaxy-wide star formation, this suggests that alongside large distortions to the global continuum SED, there is also a likely distortion to the CO SLED. 

\item  The \co emission is a factor 3 less magnified than NLR component as presented in {\bf D13a}, and factor 11 less magnified than the VLBI radio core presented in {\bf D13c}. If we assume the higher-{\sl J} CO emission regions are associated with the AGN and not galaxy-wide star formation, this would imply that the high-{\sl J} emission regions have smaller solid angles than the lower temperature gas, which is consistent with the data. This would then suggest that alongside large distortions to the global continuum SED, there is also a likely distortion to the CO SLED since high-{\sl J} CO emission regions would undergo larger amplification boosts than the low-{\sl J} emission regions. Since IRAS~10214 is often used as an archetype ULIRG, we suggest that it can not be treated as such until the spectral distortion (i.e. the function $\mu(\nu)$) has been fully characterized.  Furthermore, we note that this is a potential source of bias in the average gas properties of large samples of gravitationally lensed galaxies if not appropriately accounted for.

\item The \co emission can plausibly be divided into two components: a compact ($r_{\rm eff,app} < 3$~kpc) component associated with the higher-{\sl J} lines based on the LVG modelling and \co and \coiii spatial constraints; and an extended, lower temperature component co-spatial with the \ci\ emission. To reconcile the factor $\sim$20 difference in observed and predicted \co apparent solid angle requires the extended gas component to have a low filling factor ($\sim$0.05). If this is not the case, (i.e. filling factor $>>$0.05), it would imply that the extended gas component makes up a substantial fraction of the total molecular gas mass ($\sim$50 percent if only the relative \co flux densities of compact and extended components are considered). This fraction may be even higher than the \co flux density ratio of the two `components' implies, since a different $L_{\rm CO}$-to-$M_{\rm gas}$ conversion ratio may be appropriate for the lower pressure, low metallicity environment towards the outskirts of the gaseous disk, as has been recently shown in the local Universe \citep{Sandstrom2012}. This estimate of the gas mass in the extended CO component relative to the compact component assumes they have equal filling factors which is unlikely in reality. The literature suggests that extended, more diffuse gas components in ULIRGs have larger filling factors when compared to higher density central regions, as discussed in Sec.~\ref{sec:fillingfactor}. Disregarding opacity effects, the relative gas mass in each component should be unchanged by their relative gas filling factors. A clearer view of the impact of variable filling factors, opacity and luminosity-to-mass ratios within individual high-redshift galaxies will require the unparalleled sensitivity and angular resolution of the Atacama Large Millimeter/sub-millimeter Array (\alma).

\item Finally, we propose that the smaller high-{\sl J} region is preferentially magnified by a factor $\gtrsim2$ which would imply an even larger relative fraction of the total gas mass in the extended component.

\end{enumerate}

Strong gravitational lensing provides us with our deepest views of galaxy evolution. However, there are clearly still some challenges associated with using these cosmic telescopes. \alma will be an essential tool in high resolution mapping of higher-{\sl J} lines in both lensed (and non-lensed) high-redshift galaxies, and so probe the magnification distortion of the CO SLED. This is particularly relevant to the {\sl Herschel}-discovered lenses \citep{Negrello2010} which were FIR-selected and expected to be dominated by starbursts. Disentangling the AGN from these objects is a clear niche for {\sl e-MERLIN}, \alma\ and the \evla as we enter an era where we are able to spatially separate the AGN and star formation components in large samples of high-redshift galaxies -- a crucial probe of the interaction of these two components at cosmologically significant epochs. However, competitive angular resolution and sensitivity to observe the fundamental \co line at high redshift may eventually need the {\sl SKA} \citep{Obreschkow2011}.

\section*{Acknowledgments}

This series of papers is dedicated to the memory of Steve Rawlings. We thank the referee for very useful comments that made important improvements to the paper. We thank Christian Henkel for making his LVG software available, and Hana Schumacher for helpful discussions. The National Radio Astronomy Observatory is a facility of the National Science Foundation operated under cooperative agreement by Associated Universities, Inc. This effort/activity was supported by the European Community Framework Programme 6 and 7, Square Kilometre Array Design Studies (SKADS), contract no. 011938; and PrepSKA, grant agreement no.: 212243. RPD wishes to acknowledge funding from a NRF SKA Postdoctoral Fellowship. Any opinion, findings and conclusions or recommendations expressed in this material are those of the authors and therefore the NRF and DST do not accept any liability with regard thereto. PJM acknowledges support from the Royal Society in the form of a University Research Fellowship.


\label{lastpage}

\end{document}